\documentclass[preprints,article,accept,moreauthors,10pt,a4paper]{mdpi}

\firstpage{1}
\pubvolume{xx}
\issuenum{1}
\articlenumber{5}
\pubyear{2020}
\copyrightyear{2020}
\history{Received: date; Accepted: date; Published: date}

\usepackage{amssymb}

\usepackage{bm}
\usepackage{hyperref}
\usepackage[hyphenbreaks]{breakurl}
\usepackage{braket}
\usepackage[caption=false]{subfig}
    \usepackage[normalem]{ulem}

\newcommand{\ssa}{\widehat{\boldsymbol{\sigma}}}
\newcommand{\vv}{\mathbf{v}}
\newcommand{\vth}{v_{\text{th}}}

\newcommand{\sta}{\mathrm{st}}
\newcommand{\dif}{\mathrm{d}}

\def\bal#1\eal{\begin{align}#1\end{align}}

\Title{Kinetic Theory and Memory Effects of Homogeneous Inelastic Granular Gases under Nonlinear Drag}

\Author{Alberto Meg\'ias $^{1}$\orcidA{} Andr\'es Santos $^{2}$*\orcidB{}}

\AuthorNames{Alberto Meg\'ias and Andr\'es Santos}

\address{%
$^{1}$ \quad Departamento de F\'isica, Universidad de
Extremadura, E-06006 Badajoz, Spain; albertom@unex.es\\
$^{2}$ \quad Departamento de F\'isica and Instituto de Computaci\'on Cient\'ifica Avanzada (ICCAEx), Universidad de
Extremadura, E-06006 Badajoz, Spain; andres@unex.es}

\corres{Correspondence: andres@unex.es}

\abstract{We study a dilute granular gas immersed in a thermal bath made of smaller particles with masses not much smaller than the granular ones  in this work. Granular particles are assumed to have inelastic and hard interactions, losing energy in collisions as accounted by a constant coefficient of normal restitution. The interaction with the thermal bath is modeled by a nonlinear drag force plus a white-noise stochastic force.
The kinetic theory for this system is described by an Enskog--Fokker--Planck equation for the one-particle velocity distribution function.  To get explicit results of the temperature aging and steady states, Maxwellian and first Sonine approximations are developed. The latter takes into account the coupling of the excess kurtosis with the temperature. Theoretical predictions are  compared with direct simulation Monte Carlo and event-driven molecular dynamics simulations. While good results for the granular temperature are obtained from the Maxwellian approximation, a much better agreement, especially as inelasticity and drag nonlinearity increase,   is found when using the first Sonine approximation. The latter approximation is, additionally, crucial to account for  memory effects such as Mpemba and Kovacs-like ones.
}

\keyword{granular gases; kinetic theory; Enskog--Fokker--Planck equation; direct simulation Monte Carlo; event-driven molecular dynamics}

\begin{document}



\section{Introduction}\label{sec:1}

Since the late 20$^\text{th}$ century, the study of granular materials has become of great importance in different branches of science, such as physics, engineering, chemistry, and mathematics, motivated by either fundamental or industrial reasons. It is well known that rapid  flows in granular gases in the dilute regime are well described by a modified version of the classical Boltzmann's kinetic theory for hard particles. The most widely used model for the granular particles is the inelastic hard-sphere (IHS) one, in which particles are assumed to be hard spheres (or, generally, hard $d$-spheres) that lose energy due to inelasticity, as parameterized by a constant coefficient of normal restitution.

Theoretical predictions have been tested by different experimental setups in the freely evolving case \cite{TMHS09,YSS20}. However, it is rather difficult to  experimentally replicate the latter granular gaseous systems due to the fast freezing implied by the dissipative interactions. Then, energy  injection is very common in granular experiments \cite{PNW97,TB98,MTKSB02,HYCMW04,SGS05,AD06,EMAGWL10,MSGE22}. In addition, granular systems are never found in a vacuum on Earth. From a quick but attentive glance at our close environment, grains might be found, for example, in the form of dust or pollen suspended in the air, sand, or dirtiness, diving down or browsing through a river, or even forming part of more complex systems such as soils. Therefore, fundamental knowledge about driven granular flows contributes to the understanding of a great variety of phenomena in nature. This is one of the reasons why the study of driven granular flows has become quite important, besides its intrinsic interest at  physical and mathematical levels. Consequently, modeling driven granular flows constitutes a solid part of granular matter research, with theorists combining different collisional models and distinct interactions with the surroundings \cite{vNE98,MS00,GCV13,VS15,BBMG15,GBS18,SM09,G19,MS19,GG22}.

Recent works \cite{SP20,PSP21,MSP22} introduced a model for a molecular gas in which the interaction of the particles with a background fluid is described by a stochastic force and a drag force whose associated drag coefficient has a quadratic dependence on the velocity modulus. This latter dependence is motivated by situations where the  particle masses in the gas and the background fluid are not disparate \cite{F07,F14,HKLMSLW17}. The nonlinearity of the drag force implies an explicit coupling of the temperature with higher-order moments of the velocity distribution function (VDF) of the gas, implying the existence of interesting memory effects, such as Mpemba or Kovacs-like ones, as well as  nonexponential relaxations \cite{SP20,PSP21,MSP22}. On the other hand, the elastic property of the molecular particles implies that the system ends in an equilibrium state described by the common Maxwell--Boltzmann VDF, unlike granular gases, both driven and freely evolving \cite{vNE98,BP00,MS00,BP04,SM09,VSK14,VS15,G19,MS20}, where a coupling of the hydrodynamic quantities with the cumulants of the VDF is always present. To imagine a real situation, one might possibly consider, for example, a microgravity experiment of pollen grains in a dust cloud.

Throughout this work, we study the properties of homogeneous states of a dilute inelastic granular gas immersed in a background fluid made of smaller particles, the influence of the latter on the former being accounted for at a coarse-grained level by the sum of a deterministic nonlinear drag force and a stochastic force. This gives rise to a competition between the pure effects of the bath and the granular energy dissipation. In fact, we look into expected nonGaussianities from a Sonine approximation of the VDF, commonly used in granular gases. The theoretical results are tested against computer simulations, with special attention on the steady-state properties and memory effects.

The paper is organized as follows. We introduce the model for this system and the associated kinetic-theory evolution equations in Section~\ref{sec:2}. In Section \ref{sec:3}, the Maxwellian and first Sonine approximations are constructed, and the steady-state values are theoretically evaluated. Then, Section~\ref{sec:4} collects simulation results from the direct simulation Monte Carlo (DSMC) method and the event-driven molecular dynamics (EDMD) algorithm, which are compared to the theoretical predictions for steady and transient states, including  memory effects. Finally, some conclusions of this work are exposed in Section~\ref{sec:5}.

\section{The Model}\label{sec:2}

\begin{figure}[h!]
    \centering
    \includegraphics[width=0.8\textwidth]{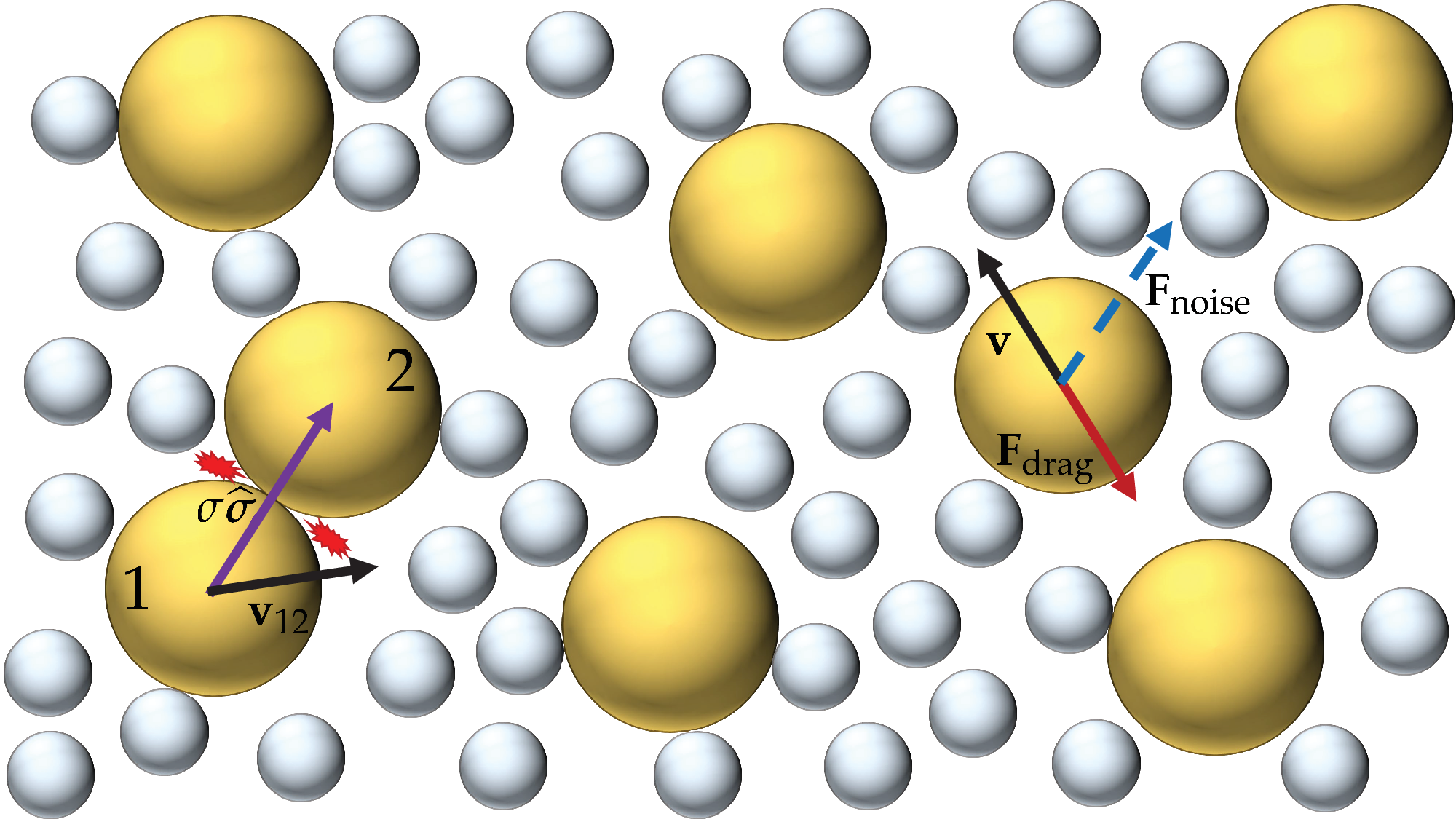}
    \caption{Illustration of the system considered in this paper.
A granular gas of hard particles (represented by large yellowish spheres) is coupled to a thermal bath (made of particles represented by the small grayish spheres) via a drag force $\mathbf{F}_{\text{drag}} = -m\xi(v)\mathbf{v}$, where $\zeta(v)$ is a velocity-dependent drag coefficient, and a stochastic force $\mathbf{F}_{\text{noise}} = m\chi(v)\boldsymbol{\eta}$, where $\boldsymbol{\eta}$ is a Gaussian white-noise term. In addition, the granular particles are subjected to binary inelastic collisions, represented by the red gleam-like lines.}
    \label{fig:sketch}
\end{figure}

We consider a homogeneous, monodisperse, and dilute granular gas of identical inelastic hard $d$-spheres of mass $m$ and diameter $\sigma$, immersed in a background fluid made of smaller particles.
In a coarse-grained description, the interactions between the grains and the fluid particles can be effectively modeled by a drag force plus a stochastic force acting on the grains.
If the mass ratio between the fluid and granular particles is not very small, the drag force becomes a nonlinear function of the velocity \cite{F07,F14,HKLMSLW17}. The model, as said in Section~\ref{sec:1}, has previously been studied in the case of elastic collisions \cite{SP20,PSP21,MSP22} but not, to our knowledge, in the context of the  IHS model. Figure \ref{fig:sketch} shows an illustration of the system and its modeling.

\subsection{Enskog--Fokker--Planck Equation}

The full dynamics of the system can be studied from the inelastic homogeneous Enskog--Fokker--Planck equation (EFPE),
\begin{align}\label{eq:EFPE}
    \partial_t f(\vv;t) -\partial_{\vv}\left[\xi(v)\vv+\frac{\chi^2(v)}{2}\partial_{\vv} \right]f(\vv;t)=J[\vv|f,f],
\end{align}
where $f$ is the one-particle VDF, so that $n = \int\dif\vv f(\vv;t)$ is the number density, and $J[\vv|f,f]$ is the usual Enskog--Boltzmann collision operator defined by
\begin{align}
\label{eq:J}
    J[\vv_1|f,f]\equiv \sigma^{d-1} g_c \int \dif\vv_2\int_{+} \dif\ssa \,(\vv_{12}\cdot\ssa) \left[\alpha^{-2}f(\vv_1^{\prime\prime})f(\vv_2^{\prime\prime})-f(\vv_1)f(\vv_2) \right].
\end{align}
Here, $\alpha$ is the coefficient of normal restitution (see below), $\vv_{12}=\vv_1-\vv_2$ is the relative velocity, $\ssa= (\mathbf{r}_1-\mathbf{r}_2)/\sigma$ is the intercenter unit vector at contact, $g_c=\lim_{r\rightarrow \sigma^{+}} g(r)$ is the contact value of the pair correlation function $g(r)$, $\int_{+} \dif\ssa \equiv \int \dif\ssa \, \Theta(\vv_{12}\cdot\ssa)$, $\Theta$ being the Heaviside step-function and $\vv_i^{\prime\prime}$ refers to the  precollisional velocity of the particle $i$. Within the IHS model, the collisional rules are expressed by \cite{G19,MS20}
\begin{equation}\label{eq:coll_rule}
    \vv_{1/2}^{\prime\prime} = \vv_{1/2} \mp \frac{1+\alpha^{-1}}{2}(\vv_{12}\cdot\ssa)\ssa.
\end{equation}
From Equation~\eqref{eq:coll_rule}, one gets $(\vv_{12}\cdot\ssa)=-\alpha (\vv_{12}^{\prime\prime}\cdot\ssa)$; this relation defines the coefficient of normal restitution, which is assumed to be constant.

The second term on the left-hand side of Equation~\eqref{eq:EFPE} represents the action of a net force $\mathbf{F}=\mathbf{F}_{\text{drag}}+\mathbf{F}_{\text{noise}}$ describing the interaction with the particles of the background fluid. The deterministic nonlinear drag force is $\mathbf{F}_{\text{drag}}=-m\xi(v)\mathbf{v}$, where the drag coefficient $\xi(v)$ depends on the velocity. In turn, $\mathbf{F}_{\text{noise}}=m\chi^2(v)\boldsymbol{\eta}$ is a stochastic force, where $\chi^2(v)$ measures its intensity, and $\boldsymbol{\eta}$ is a stochastic vector with the properties of a zero-mean Gaussian white noise with a unit covariance matrix, i.e.,
\begin{equation}
    \langle \boldsymbol{\eta}_i(t)\rangle = 0, \quad \langle \boldsymbol{\eta}_i(t) \boldsymbol{\eta}_j(t^\prime) \rangle = \mathsf{I}\delta_{ij} \delta(t-t^\prime),
\end{equation}
where $i$ and $j$ are particle indices, and $\mathsf{I}$ is the $d\times d$ unit matrix so that different Cartesian components of $\boldsymbol{\eta}_i(t)$ are uncorrelated. The functions $\xi(v)$ and $\chi^2(v)$ are constrained to follow the fluctuation-dissipation theorem as
\begin{equation}
 \chi^2(v) = v^2_b \xi(v),
\end{equation}
 $v_b = \sqrt{2T_b/m}$ being the thermal velocity associated with the background temperature $T_b$.

The drag coefficient $\xi$ is commonly assumed to be independent of the velocity. However, a dependence on $v$ cannot be ignored if the mass of a fluid particle is not much smaller than that of grain \cite{F07,F14,HKLMSLW17}. The first correction to $\xi=\text{const}$ is a quadratic term \cite{SP20,PSP21,MSP22}, namely
\begin{equation}
   \xi(v)= \xi_0\left(1+2\gamma \frac{v^2}{v_b^2}\right),
\end{equation}
where $\xi_0$ is the drag coefficient in the zero-velocity limit and $\gamma$ controls  the degree of nonlinearity of the drag force.

\subsection{Dynamics}

It is well known that, in the case of driven granular gases \cite{vNE98,MS00,SM09,CVG12,CVG13,VS15,G19,MS19}, there exists a competition between the loss and gain of energy due to inelasticity and the action of the thermal bath, respectively. This eventually leads the granular gas to a steady state, in contrast to the freely cooling case \cite{G19}.

The basic macroscopic quantity characterizing the time evolution of the system is the granular temperature, defined analogously to the standard temperature in kinetic theory as
\begin{equation}
    T(t) = \frac{m}{dn}\int \dif\vv \, v^2 f(\vv; t).
\end{equation}
While in the case of elastic collisions, the asymptotic steady state is that of equilibrium at temperature $T_b$, i.e., $\lim_{t\to\infty}T(t)=T_b$, in the IHS model, the steady state is a nonequilibrium one and, moreover,  $\lim_{t\to\infty}T(t)=T^\sta<T_b$.
From  the EFPE, one can derive the evolution equation of the granular temperature, which is given by
\begin{equation}\label{eq:T_ev}
 \frac{\partial_t T}{\xi_0} =2(T_b-T)\left[1+(d+2)\gamma \frac{T}{T_b}\right]-2(d+2)\gamma \frac{T^2}{T_b}a_2-\frac{\zeta}{\xi_0} T,
\end{equation}
where
\begin{equation}
\label{eq:zeta}
    \zeta(t) \equiv  -\frac{m}{d T(t) n} \int \dif\vv\, v^2 J[\vv,f,f]
\end{equation}
is the cooling rate and
\begin{equation}
    a_2(t) \equiv \frac{d}{d+2}\frac{n\int \dif\vv \, v^4 f(\vv;t)}{\left[\int \dif\vv \, v^2 f(\vv;t)\right]^2}-1
\end{equation}
is the excess kurtosis (or fourth cumulant) of the time-dependent VDF.
The coupling of $T(t)$ to $a_2(t)$ is a direct consequence of the quadratic term in the drag coefficient. As for the cooling rate $\zeta(t)$, it is a consequence of inelasticity and, therefore, vanishes in the elastic case (conservation of energy). Insertion of Equation~\eqref{eq:J} into Equation~\eqref{eq:zeta} yields \cite{G19}
\begin{equation}
\label{eq:zeta_bis}
    \zeta(t) =(1-\alpha^2)\frac{\nu(t)}{\sqrt{2}dn^2}\frac{\Gamma\left(\frac{d}{2}\right)}{\Gamma\left(\frac{d+3}{2}\right)}\int \dif\vv_1 \int\dif\vv_2 \,\left[\frac{v_{12}}{\vth(t)}\right]^3 f(\vv_1;t)f(\vv_2;t).
\end{equation}
Here, $\vth(t)=\sqrt{2T(t)/m}$ is the time-dependent thermal velocity and
\begin{equation}
\nu(t)=g_c K_d n\sigma^{d-1}\vth(t),\quad K_d\equiv \frac{\pi^{d-1}}{\sqrt{2}\Gamma(d/2)},
\end{equation}
is the time-dependent collision frequency.

Let us rewrite Equation~\eqref{eq:T_ev} in dimensionless form. First, we introduce the reduced quantities
\begin{equation}
\label{eq:dimensionless}
 t^*\equiv \nu_b t,\quad \theta(t^*)\equiv \frac{T(t)}{T_b},\quad \xi_0^*\equiv\frac{\xi_0}{\nu_b},\quad \mu_{\ell}(t^*) \equiv -\frac{1}{n\nu(t)}\int\dif\vv\medspace \left[\frac{v}{\vth(t)}\right]^\ell {J}[\vv|f,f],
\end{equation}
where $\nu_b=g_c K_d n\sigma^{d-1}v_b$ is the collision frequency associated with the background temperature $T_b$.
Note that the control parameter $\xi_0^*$ measures the ratio between the characteristic times associated with collisions and drag. In the molecular case, $\xi_0^*$ depends on the bath-to-grain density, size, and mass ratios, but otherwise, it is independent of $T_b$ \cite{HKLMSLW17,SP20}. In terms of the quantities defined in Equation~\eqref{eq:dimensionless}, Equation~\eqref{eq:T_ev} becomes
\begin{equation}
\label{eq:theta_ev}
    \frac{\dot{\theta}}{\xi_0^*}=2(1-\theta)\left[1+(d+2)\gamma\theta\right]-2(d+2)\gamma\theta^2a_2-\frac{2 \mu_2}{d}\frac{\theta^{3/2}}{\xi_0^*},
\end{equation}
where henceforth, a dot over a quantity denotes a derivative with respect to $t^*$, and we have taken into account that $\zeta(t)/\nu(t)=2\mu_2(t^*)/d$ and $\nu(t)/\nu_b=\theta^{1/2}(t^*)$.

Equation \eqref{eq:theta_ev} is not a closed equation since it is coupled to the full VDF through $a_2$ and $\mu_2$. More  generally, taking velocity moments on  the EFPE, an infinite hierarchy of moment equations can be derived.
In dimensionless form, it reads
\begin{align}
\label{eq:hierarchy_M}
   \frac{\dot{M_\ell}}{\xi^*_0}=&\ell\left\{ \left[(\ell-2)\gamma+\frac{\mu_2}{d}\frac{\sqrt{\theta}}{\xi_0^*}+(d+2)\gamma\theta(1+a_2)-\frac{1}{\theta}\right]M_\ell-2 \gamma\theta M_{\ell+2}+\frac{d+\ell-2}{2}\frac{M_{\ell-2}}{\theta}\right\}
       \nonumber \\ &-\mu_\ell\frac{\sqrt{\theta}}{\xi_0^*},
\end{align}
where $M_\ell(t^*) \equiv n^{-1}\int\dif\vv\, [v/\vth(t)]^\ell f(\vv;t)$. In particular, $M_0=1$, $M_2=\frac{d}{2}$, $M_4=\frac{d(d+2)}{4}(1+a_2)$, and $M_6=\frac{d(d+2)(d+4)}{8}(1+3a_2-a_3)$, $a_3$ being the sixth cumulant.

Equation \eqref{eq:hierarchy_M} is trivial for $\ell=0$ and $\ell=2$. The choice $\ell=4$ yields
\begin{align}
    \frac{\dot{a_2}}{\xi_0^*}=&4\gamma\theta\left[ \frac{2(1+a_2)}{\theta}+(d+2)(1+a_2)^2-(d+4)(1+3a_2-a_3)\right]-4\frac{a_2}{\theta}
    \nonumber \\
&+\frac{4}{d}\left[\mu_2(1+a_2)
        -\frac{\mu_4}{d+2}\right]\frac{\sqrt{\theta}}{\xi^*_0}.\label{eq:a2_ev}
\end{align}

Equations \eqref{eq:theta_ev}--\eqref{eq:a2_ev}  are formally exact in the context of the EFPE, Equation~\eqref{eq:EFPE}. Nevertheless, they cannot be solved  because of the infinite nature of the hierarchy \eqref{eq:hierarchy_M} and the highly nonlinear dependence of the collisional moments $\mu_\ell$ on the velocity moments of the VDF. This forces us to devise tractable approximations in order to extract information about the dynamics and steady state of the system.

\section{Approximate Schemes}\label{sec:3}

\subsection{Maxwellian Approximation}\label{sec:3.1}

The simplest approximation consists of assuming that  the VDF remains very close to a Maxwellian during its time evolution so that the excess kurtosis $a_2$ can be neglected in Equation~\eqref{eq:theta_ev}, and the reduced cooling rate $\mu_2$ can be approximated by \cite{GS95,vNE98,MS00,BP04,BP06,SM09,G19}
\begin{equation}
    \mu_2 \approx \mu_2^{(0)} = 1-\alpha^2.
\end{equation}
In this Maxwellian approximation (MA), Equation~\eqref{eq:theta_ev} becomes
\begin{equation}
\label{eq:theta_ev_Max}
    \frac{\dot{\theta}}{\xi_0^*}\approx 2(1-\theta)\left[1+(d+2)\gamma\theta\right]-\frac{2(1-\alpha^2)}{d}\frac{\theta^{3/2}}{\xi_0^*}.
\end{equation}
This is a closed equation for the temperature ratio $\theta(t^*)$ that can be solved numerically for any initial temperature. The steady-state value $\theta^\sta$ in the MA is obtained by equating to zero the right-hand side of Equation~\eqref{eq:theta_ev_Max}, which results in a fourth-degree algebraic equation.

\subsection{First Sonine Approximation}\label{sec:3.2}

As we will see later, the MA given by Equation~\eqref{eq:theta_ev_Max} provides a simple and, in general, rather accurate estimate of $\theta(t^*)$ and $\theta^\sta$. However, since the evolution of temperature is governed by its initial value only, the MA is unable to capture  memory phenomena, such as Mpemba- or Kovacs-like effects, which are  observed even in the case of elastic particles \cite{SP20,PSP21,MSP22}. This is a consequence of the absence of any coupling of $\theta$ with some  other dynamical variable(s).

The next simplest approximation beyond the MA consists of incorporating $a_2$ into the description but assuming it is small enough as to neglect nonlinear terms involving this quantity, as well as higher-order cumulants, i.e., $a_2^k\to 0$ for $k\geq 2$ and $a_\ell\to 0$ for $\ell\geq 3$.
This represents the so-called first Sonine approximation (FSA), according to which Equations~\eqref{eq:theta_ev} and \eqref{eq:a2_ev} become
\begin{subequations}
\label{eq:ev_FSA}
\begin{equation}
\frac{\dot{\theta}}{\xi_0^*}\approx 2(1-\theta)\left[1+(d+2)\gamma\theta\right]-2(d+2)\gamma\theta^2a_2-\frac{2\left[ \mu_2^{(0)}+\mu_2^{(1)}a_2\right]}{d}\frac{\theta^{3/2}}{\xi_0^*},
\label{eq:theta_ev_FirstSonine}
\end{equation}
\begin{align}
    \frac{\dot{a_2}}{\xi_0^*}\approx&4\gamma\theta\left[ 2\frac{1+a_2}{\theta}+(d+2)(1+2a_2)-(d+4)(1+3a_2)\right]-4\frac{a_2}{\theta}
      \nonumber \\
   & +\frac{4}{d}\left\{\mu_2^{(0)}-\frac{\mu_4^{(0)}}{d+2}+\left[\mu_2^{(0)}+ \mu_2^{(1)}-\frac{\mu_4^{(1)}}{d+2}\right]a_2\right\}\frac{\sqrt{\theta}}{\xi^*_0}
   ,
    \label{eq:a2_ev_FirstSonine}
\end{align}
\end{subequations}
where we have used \cite{GS95,vNE98,MS00,BP04,BP06,SM09,G19}
\begin{align}\label{eq:moments_approx}
    \mu_2 \approx \mu_2^{(0)}+ \mu_2^{(1)}a_2, \quad \mu_4 \approx \mu_4^{(0)}+ \mu_4^{(1)}a_2,
\end{align}
with
\begin{subequations}
\label{eq:moments_approx_expr}
\begin{align}
    \mu_2^{(1)} = \frac{3}{16}\mu_2^{(0)}, \quad \mu_4^{(0)} = \left(d+\frac{3}{2}+\alpha^2\right)\mu_2^{(0)}, \\
    \mu_4^{(1)} = \frac{3}{32}\left(10d+39+10\alpha^2 \right)\mu_2^{(0)}+(d-1)(1+\alpha).
\end{align}
\end{subequations}
Equations \eqref{eq:ev_FSA} make a set of two coupled differential equations. In contrast to the MA, now the evolution of $\theta(t^*)$ is governed by the initial values of both $\theta$ and $a_2$. This latter fact implies that the evolution of temperature depends on the initial preparation of the whole VDF, this being a determinant condition for the emergence of memory effects, which will be explored later in Section~\ref{sec:4.1}.

\subsubsection{Steady-State Values}
The steady-state values $\theta^\sta$ and $a_2^\sta$ in the FSA are obtained by equating to zero the right-hand sides of Equations~\eqref{eq:ev_FSA}, i.e.,
\begin{subequations}
\label{eq:eqs_AI}
\begin{align}
\label{eq:eqs_AI_1}
    \dot{\theta}=0\Rightarrow F_0(\theta^\sta) + F_1(\theta^\sta)a_2^\sta =& \left[\mu_2^{(0)}+\mu_2^{(1)}a_2^\sta \right]\frac{(\theta^\sta)^{3/2}}{\xi^*_0},\\
    \label{eq:eqs_AI_2}
    \dot{a_2}=0\Rightarrow G_0(\theta^\sta)+ G_1(\theta^\sta)a_2^\sta =&\left\{\frac{\mu_4^{(0)}}{d+2}-\mu_2^{(0)}+\left[\frac{\mu_4^{(1)}}{d+2}-\mu_2^{(0)}-\mu_2^{(1)}\right] a_2^\sta\right\}\frac{(\theta^\sta)^{3/2}}{\xi^*_0},
\end{align}
\end{subequations}
where
\begin{subequations}
\label{eq:F0-G1}
\begin{align}
    F_0(\theta)=&d(1-\theta)\left[1+(d+2)\gamma\theta\right], \quad F_1(\theta) = -d(d+2)\gamma\theta^2, \\
    G_0(\theta) =& 2d\gamma \theta(1-\theta), \quad G_1(\theta) = d\gamma\theta\left[2-\theta(d+8)\right]-d.
\end{align}
\end{subequations}
Eliminating $a_2^\sta$ in Equation~\eqref{eq:eqs_AI}, one gets a closed nonlinear equation for $\theta^\sta$ in our FSA. Once numerically solved, $a_2^\sta$ is simply given by either Equation~\eqref{eq:eqs_AI_1} or Equation~\eqref{eq:eqs_AI_2}. For instance, Equation~\eqref{eq:eqs_AI_1} gives
\begin{equation}\label{eq:a2st}
    a_2^\sta = -\frac{F_0(\theta^\sta)-\mu_2^{(0)}(\theta^\sta)^{3/2}/\xi_0^*}{F_1(\theta^\sta)-\mu_2^{(1)}(\theta^\sta)^{3/2}/\xi_0^*}.
\end{equation}

\begin{figure}[H]
   \centering
    \includegraphics[width=0.43\textwidth]{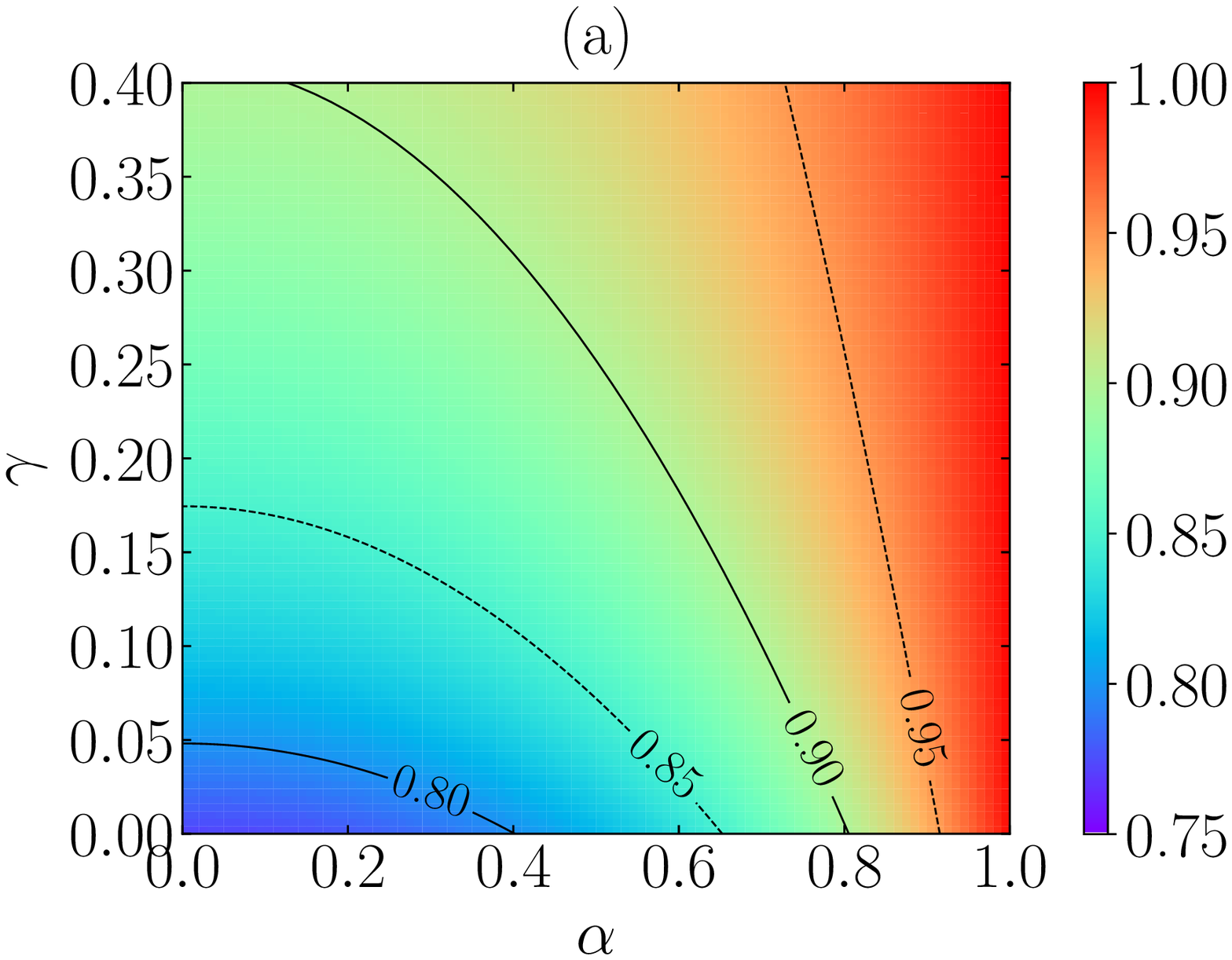}
    \includegraphics[width=0.43\textwidth]{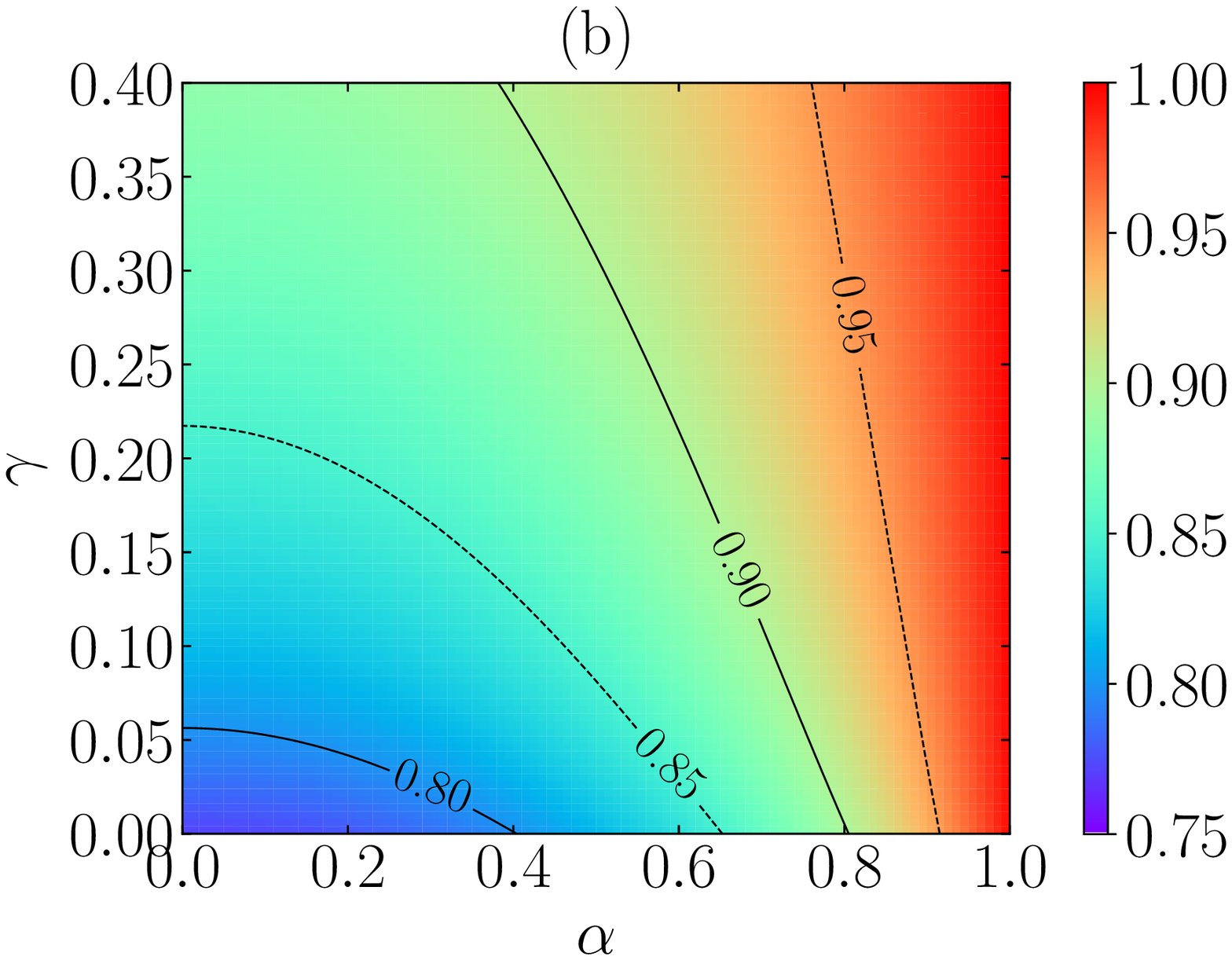}\\
    \includegraphics[width=0.43\textwidth]{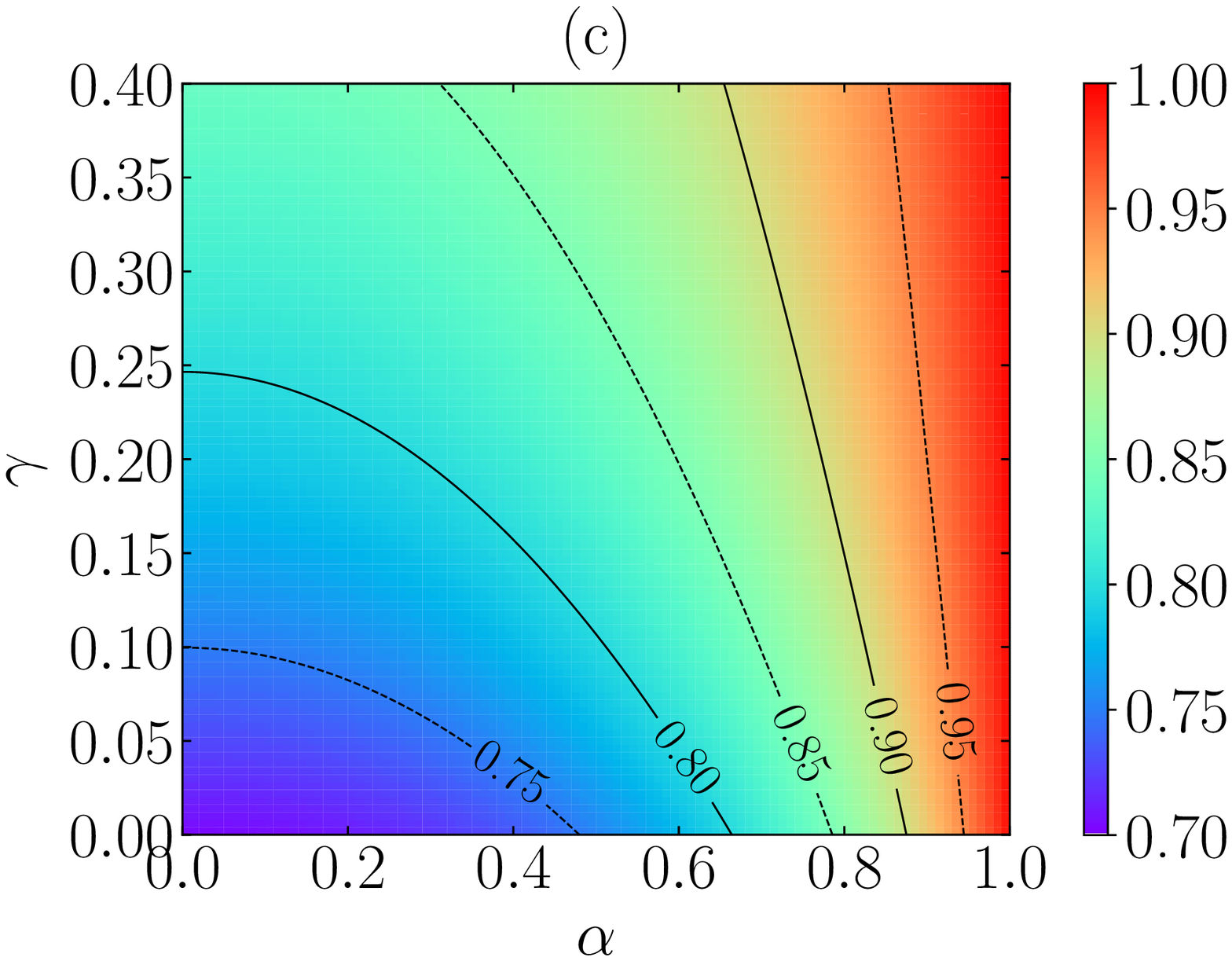}
    \includegraphics[width=0.43\textwidth]{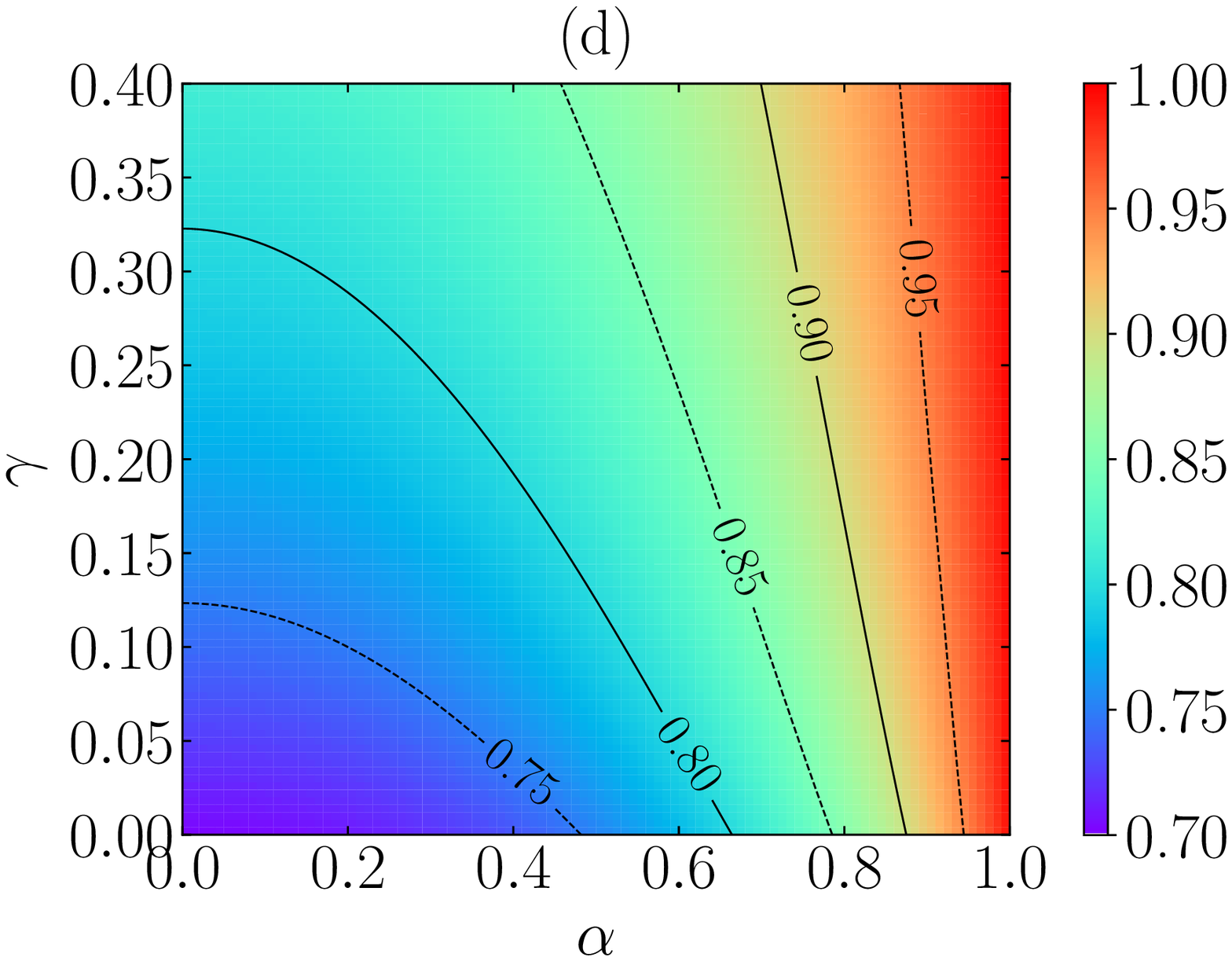}
    \caption{Theoretical predictions for the steady-state value of the reduced temperature $\theta^\sta$ as a function of the coefficient of normal restitution $\alpha$ and of the nonlinearity control parameter $\gamma$ with $\xi_0^*=1$.  Panels (\textbf{a},\textbf{c}) correspond to the MA, while panels (\textbf{b},\textbf{d}) correspond to the FSA. The dimensionality of the system is $d=3$ in panels (\textbf{a},\textbf{b}) and $d=2$ in panels (\textbf{c},\textbf{d}). The contour lines are separated by an amount of $\Delta \theta^\sta=0.05$.}
    \label{fig:thetast_th}
\end{figure}

Figure \ref{fig:thetast_th} compares the MA and FSA predictions of $\theta^\sta$ for three- and two-dimensional granular gases with $\xi_0^*=1$. We observe that the breakdown of equipartition (as measured by $1-\theta^\sta$) is stronger in 2D than 3D and increases with increasing inelasticity but decreases as the nonlinearity of the drag force grows. Apart from that, the deviations of the MA values with respect to the FSA ones increase with increasing nonlinearity and inelasticity, the MA values tending to be larger (i.e., closer to equipartition) than the FSA ones.

\begin{figure}[H]
    \centering
    \includegraphics[width=0.47\textwidth]{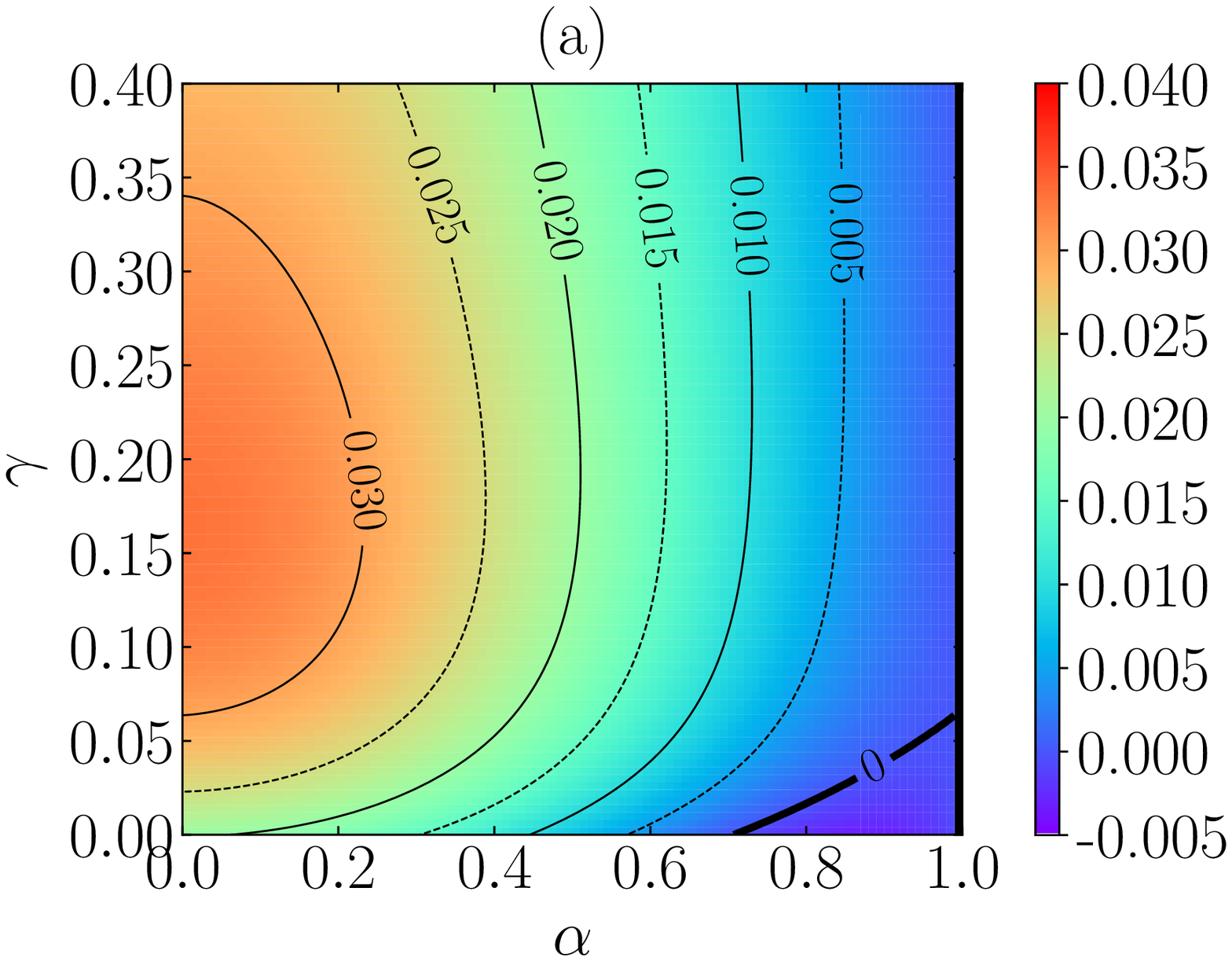}
    \includegraphics[width=0.47\textwidth]{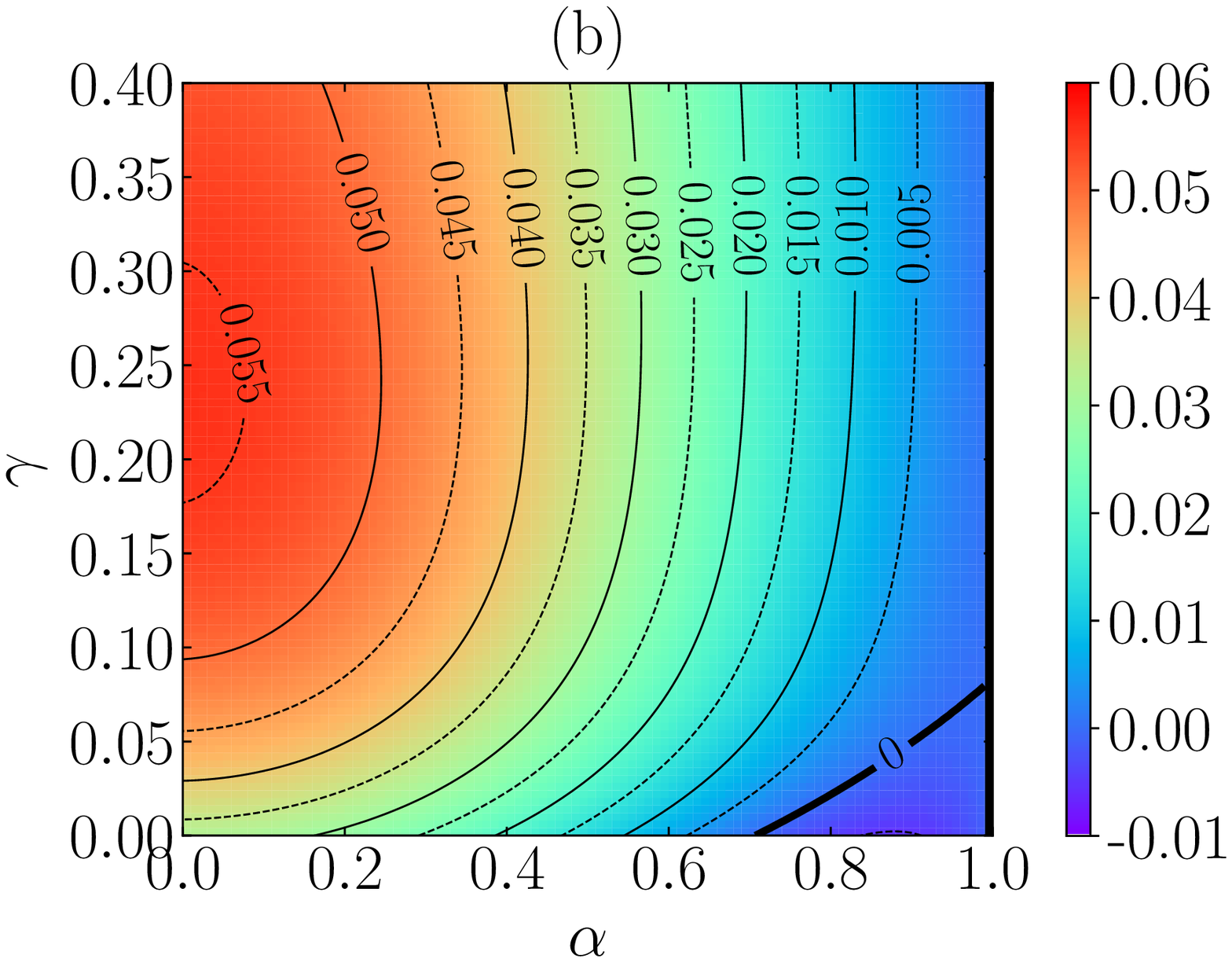}
    \caption{FSA predictions for the steady-state value of the excess kurtosis $a_2^\sta$ as a function of the coefficient of normal restitution $\alpha$ and of the nonlinearity control parameter $\gamma$ with $\xi_0^*=1$.  The dimensionality of the system is $d=3$ in panel (\textbf{a}) and $d=2$ in panel (\textbf{b}). The contour lines are separated by an amount of $\Delta a_2^\sta=0.005$. The thickest black line corresponds to the contour $a_2^{\sta}=0$.}
    \label{fig:a2st_th}
\end{figure}

The FSA predictions of $a_2^\sta$ are displayed in Figure~\ref{fig:a2st_th}. First, it is quite apparent that the departure from the Maxwellian VDF (as measured by the magnitude of $a_2^\sta$) is higher in 2D than 3D. It is also noteworthy that $a_2^\sta$ starts growing with increasing $\gamma$, reaches a maximum at a certain value $\gamma=\gamma_{\max}(\alpha,\xi_0^*)$, and then it decreases as $\gamma$ increases beyond $\gamma_{\max}(\alpha,\xi_0^*)$; this effect is more pronounced for small $\alpha$.

Another interesting feature is that $a_2^\sta$ takes negative values (in the domain of small inelasticity) only if $\gamma$ is smaller than a certain value $\gamma_c$. Of course,  $\left.a_2^\sta(\alpha,\gamma)\right|_{\alpha=1}=0$ for any $\gamma$ (since the steady state with $\alpha=1$ is that of equilibrium), but $\left.\partial_\alpha a_2^\sta(\alpha,\gamma)\right|_{\alpha=1}<0$ if $\gamma<\gamma_c$ and $\left.\partial_\alpha a_2^\sta(\alpha,\gamma)\right|_{\alpha=1}>0$ if $\gamma>\gamma_c$. Thus, the critical value $\gamma_c$ is determined by the condition
$\left.\partial_\alpha a_2^\sta(\alpha,\gamma_c)\right|_{\alpha=1}=0$. Interestingly, the result obtained from the FSA, Equation~\eqref{eq:a2st}, is quite simple, namely
\begin{equation}\label{eq:gamma_c}
    \gamma_c = \frac{1}{3(d+2)},
\end{equation}
which is independent of $\xi_0^*$.

\subsubsection{Special Limits}\label{sec:3.2.2}

\paragraph*{Absence of Drag}

Let us first define a \emph{noise temperature} $T_n$ as
$T_n = T_b \xi_0^{*2/3}\propto (\xi_0T_b)^{2/3}$, so that $\theta^{3/2}/\xi_0^*=(T/T_n)^{3/2}$.
Now we take the limit of zero drag, $\xi_0\to 0$, with finite noise temperature $T_n$. This implies $T_b\to \infty$, and thus, the natural temperature scale of the problem is no longer $T_b$ but $T_n$, i.e., $\theta^\sta\to 0$ but $T^\sta/T_n=\text{finite}$.
From Equations~\eqref{eq:F0-G1} we see that $F_0(0)= d$, $F_1(0)= 0$, $G_0(0)= 0$, and $G_1(0)= -d$. Therefore, Equations~\eqref{eq:eqs_AI} reduce to
\begin{subequations}
\label{eq:eqs_AI_bis}
\begin{align}
\label{eq:eqs_AI_bis1}
   \dot{\theta}=0\Rightarrow d\left(\frac{T_n}{T^\sta}\right)^{3/2}=& \mu_2^\sta ,\\
   \dot{a_2}=0\Rightarrow \label{eq:eqs_AI_bis2}
    -d\left(\frac{T_n}{T^\sta}\right)^{3/2}a_2^\sta =&\frac{\mu_4^\sta}{d+2}-\mu_2^\sta(1+a_2^\sta),
\end{align}
\end{subequations}
where, for the sake of generality, we have undone the linearizations with respect to $a_2^\sta$. By the elimination of $\left({T_n}/{T^\sta}\right)^{3/2}$, one simply gets
$(d+2)\mu_2^\sta=\mu_4^\sta$, from which one can then obtain $a_2^\sta$ upon linearization \cite{vNE98,MS00}. The steady-state temperature is given by $T^\sta/T_n=(d/\mu_2^\sta)^{2/3}$.

\paragraph*{Homogeneous Cooling State}
If, in addition to $\xi_0\to 0$, we take the limit $T_n\to 0$, the asymptotic state becomes the homogeneous cooling state. In that case, $T$ does not reach a true stationary value, but $a_2$ does. As a consequence, Equation~\eqref{eq:eqs_AI_bis1} is not applicable, but Equation~\eqref{eq:eqs_AI_bis2}, with $T_n=0$, can still be used to get  $(d+2)\mu_2^\sta(1+a_2^\sta)=\mu_4^\sta$, as expected \cite{vNE98,MS00,SM09}.

\paragraph*{Linear Drag Force}
If the drag force is linear in velocity (i.e., $\gamma=0$), we have $F_0(\theta)=d(1-\theta)$, $F_1(\theta)=0$, $G_0(\theta)=0$, and $G_1(\theta)=-d$. Using Equation~\eqref{eq:eqs_AI_2}, $a_2^\sta$ is given by
\begin{equation}
    a_2^\sta = -\frac{\mu_4^{(0)}-(d+2)\mu_2^{(0)}}{\mu_4^{(0)}-(d+2)\left[\mu_2^{(0)}+\mu_2^{(1)}-{d\xi_0^*}/{(\theta^\sta)^{3/2}}\right]},
\end{equation}
thus recovering previous results \cite{CVG12,CVG13}.

\paragraph*{Collisionless Gas}

If the collision frequency $\nu_b$ is much smaller than the zero-velocity drag coefficient $\xi_0$, the granular dynamics is dominated by the interaction with the background fluid and the grain--grain collisions can be neglected; therefore,  the grains behave as Brownian particles. In that case, the relevant dimensionless time is no longer $t^*=\nu_bt$ but $\tau=\xi_0 t=\xi_0^* t^*$ and the evolution equations \eqref{eq:ev_FSA} become

\begin{subequations}
\label{eq:ev_FSABr}
\begin{equation}
\frac{\dif{\theta}}{\dif\tau}\approx 2(1-\theta)\left[1+(d+2)\gamma\theta\right]-2(d+2)\gamma\theta^2a_2,
\label{eq:theta_ev_FirstSonineBr}
\end{equation}
\begin{equation}
    \frac{\dif{a_2}}{\dif \tau}\approx 4\gamma\theta\left[ 2\frac{1+a_2}{\theta}+(d+2)(1+2a_2)-(d+4)(1+3a_2)\right]-4\frac{a_2}{\theta}   ,
    \label{eq:a2_ev_FirstSonineBr}
\end{equation}
\end{subequations}
It is straightforward to check that the steady-state solution is $\theta^\sta=1$ and $a_2^\sta=0$, regardless of the value of $\gamma$, as expected.

\section{Comparison with Computer Simulations}\label{sec:4}

We have carried out DSMC and EDMD computer simulations to validate the theoretical predictions. The DSMC method is based on the acceptance-rejection Monte Carlo Metropolis decision method \cite{MRRTT53} but adapted to solve the Enskog--Boltzmann equation~\cite{B94,B13}, and the algorithm is, consequently, adjusted to agree with the inelastic collisional model \cite{MS00,SM09} and reflect the interaction with the bath \cite{MSP22}. On the other hand, the EDMD algorithm is based on the one exposed in Ref.~\cite{MSP22}, but is adequated to the IHS collisional model. The main difference between DSMC and EDMD is that the latter does not follow any statistical rule to solve the Boltzmann equation but solves the equations of motion of the hard particles. Simulation details about the characteristics of the schemes and numerical particularities can be found in Appendix~\ref{sec:ap1}.

In Figure~\ref{fig:steady_states_sim}, results from simulations are compared with the theoretical predictions of $\theta^\sta$ (from MA and FSA) and of $a_2^\sta$ (from FSA) in a three-dimensional ($d=3$) IHS system with $\xi_0^*=1$.
It can be observed that both the DSMC and EDMD results agree with each other. From Figure \ref{fig:steady_states_sim}a, one can conclude that, as expected, FSA works in the prediction of $\theta^\sta$ much better than MA for values of $\gamma$ close to $\gamma_{\max}(\alpha,\xi_0^*)$ (which corresponds to the maximum magnitude of $a_2^\sta$). Moreover, FSA gives reasonably good estimates for the values of $a_2^\sta$, although they get worse for increasing inelasticity, i.e., decreasing $\alpha$. One might also think  that the increase in $\gamma$ produces a poorer approach; however, according to the theory, the performance of FSA improves if  $\gamma>\gamma_{\max}(\alpha,\xi_0^*)$, which corresponds to a decrease in $|a_2^\sta|$. Of course, nonlinear terms or higher-order cumulants might play a role that  is not accounted for within FSA.

\begin{figure}[H]
    \centering
    \includegraphics[width=0.45\textwidth]{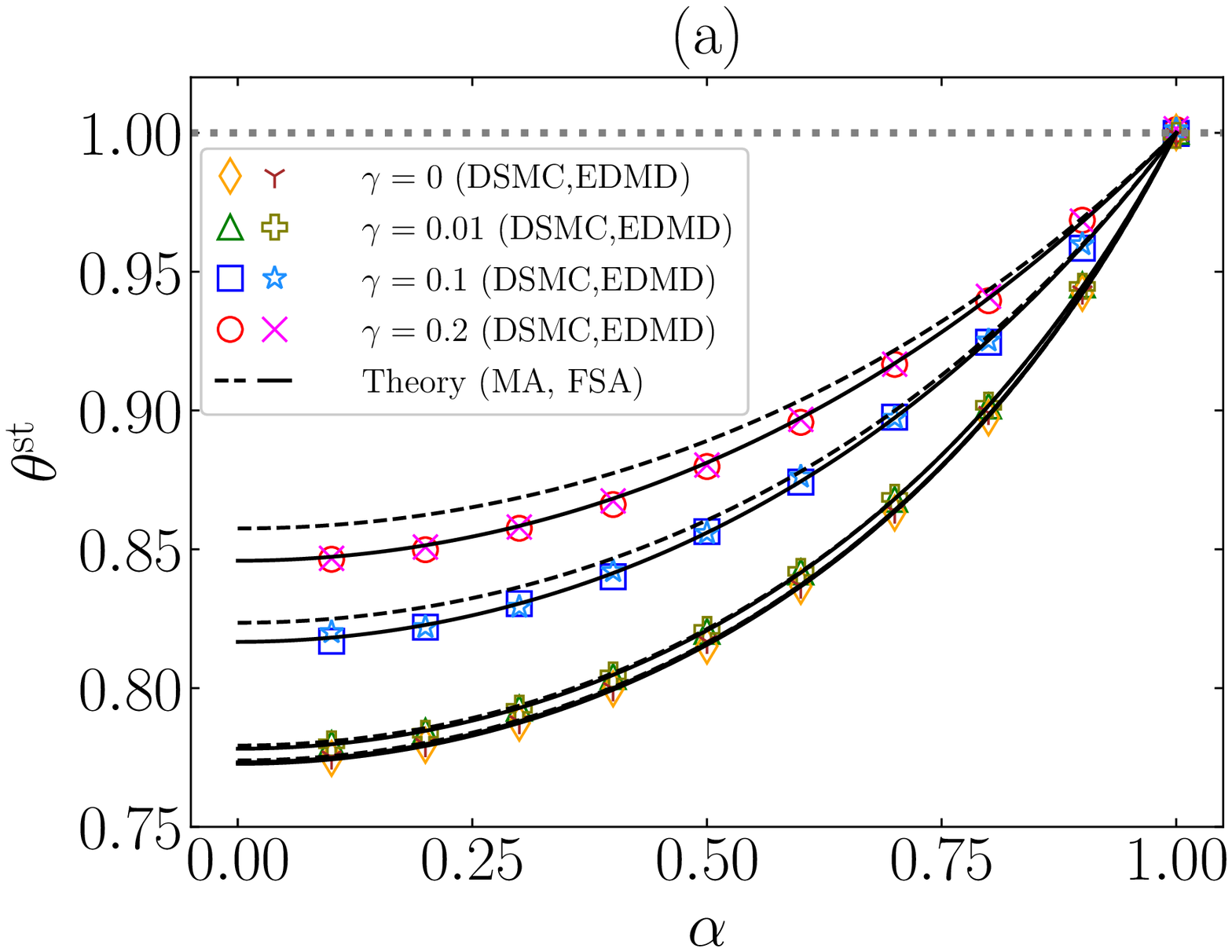}
    \includegraphics[width=0.45\textwidth]{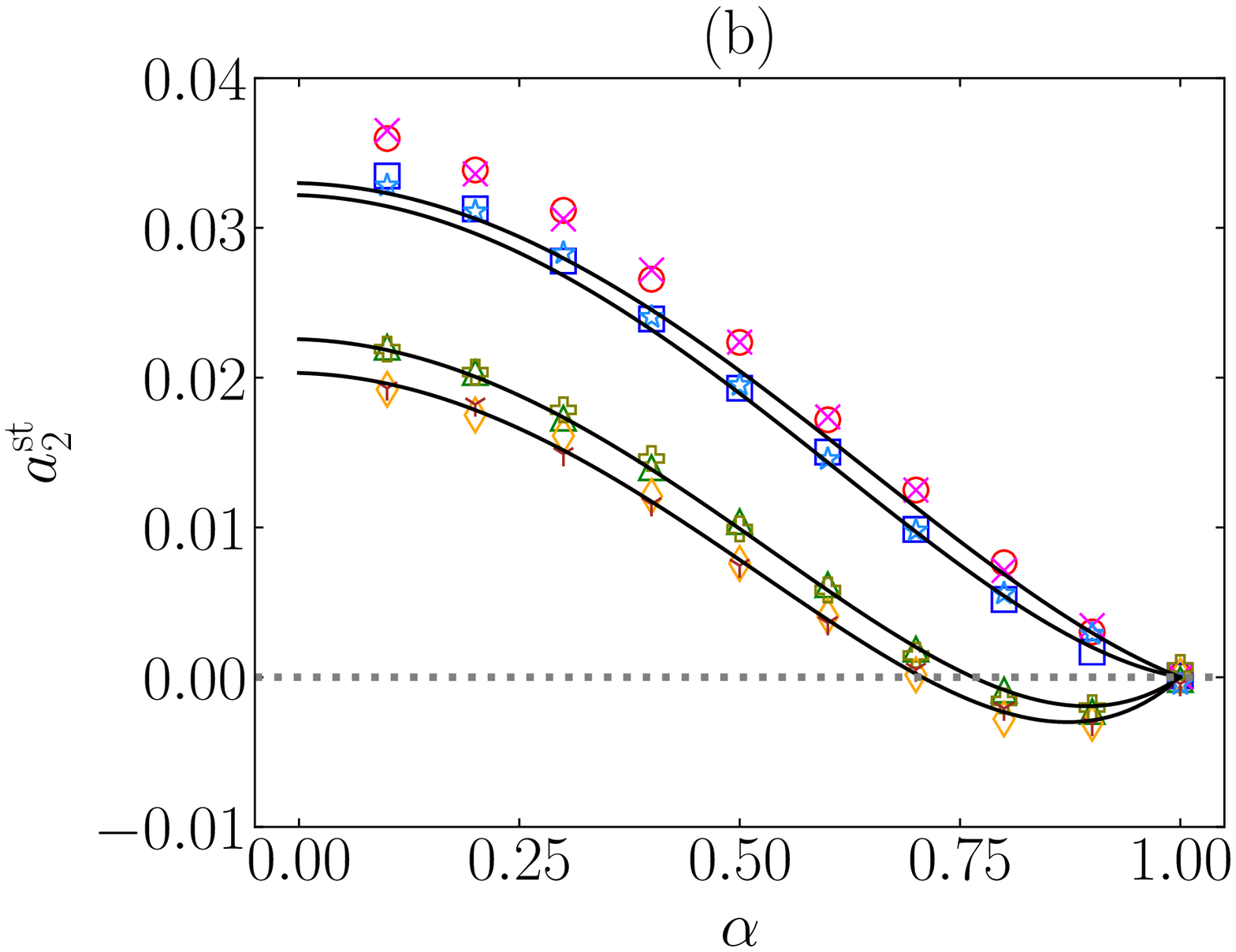}
    \caption{Plots of the steady-state values of (\textbf{a}) the temperature ratio $\theta^\sta$ and (\textbf{b}) the excess kurtosis $a_2^\sta$ vs. the coefficient of normal restitution $\alpha$ for $\xi_0^*=1$, $d=3$, and different values of the nonlinear parameter: $\gamma=0,0.01,0.1,0.2$. The symbols stand for DSMC ($\diamond$, $\triangle$, $\square$,  $\circ$) and EDMD ( $\mathsf{Y}$, $+$, $*$, $\times$) simulation results, respectively. Dashed (-- --) and solid (----) lines refer to MA (only in panel (\textbf{a})) and FSA predictions, respectively. The horizontal gray dotted lines ($\cdots$) correspond to the steady-state values in the elastic limit. As representative values, note that, at $\xi_0^*=1$, one has  $\gamma_{\max}=0.25, 0.19, 0.17$  for $\alpha=0.8, 0.5, 0.2$, respectively. }
    \label{fig:steady_states_sim}
\end{figure}

\begin{figure}[H]
    \centering
    \includegraphics[width=0.245\textwidth]{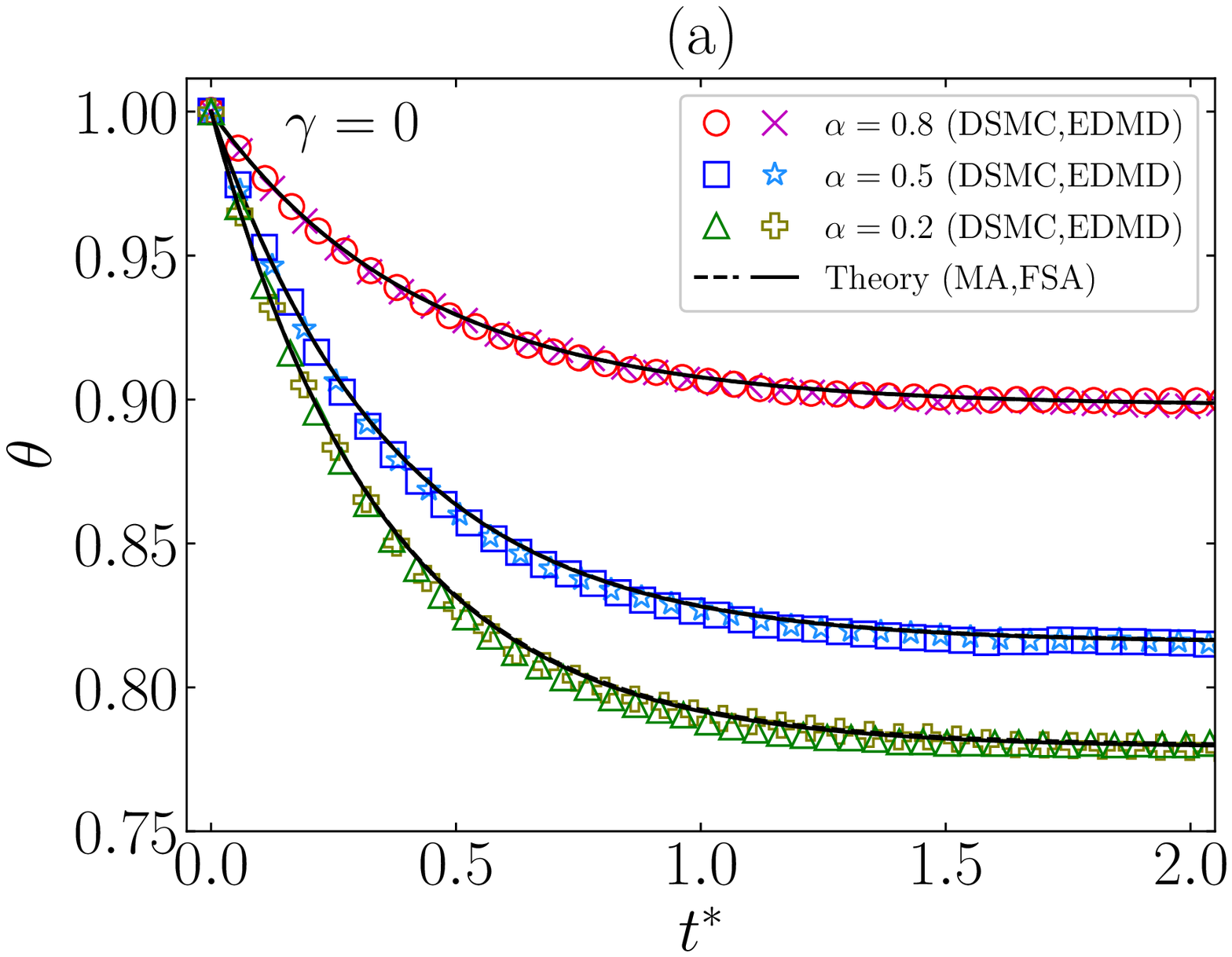}
    \includegraphics[width=0.245\textwidth]{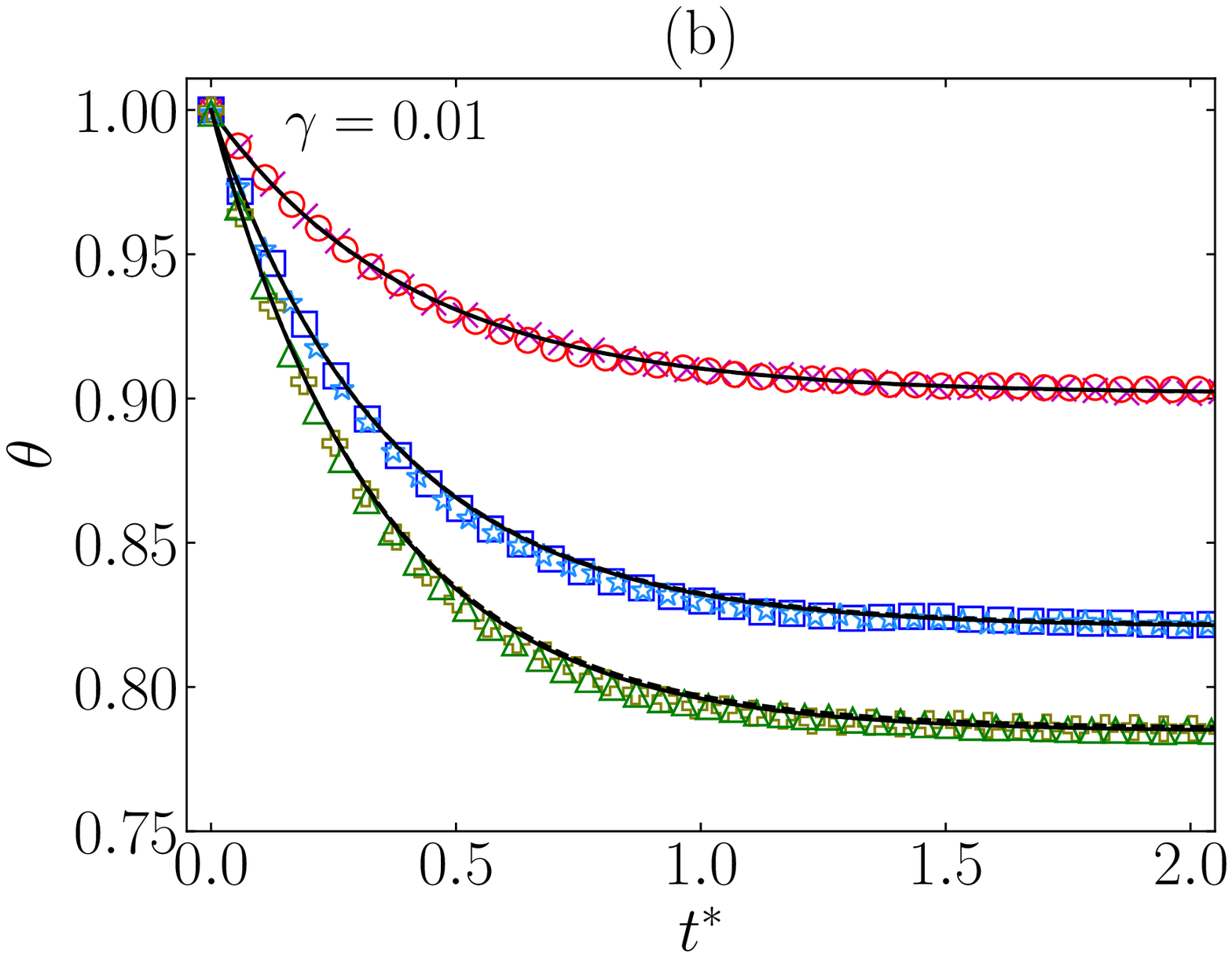}
    \includegraphics[width=0.245\textwidth]{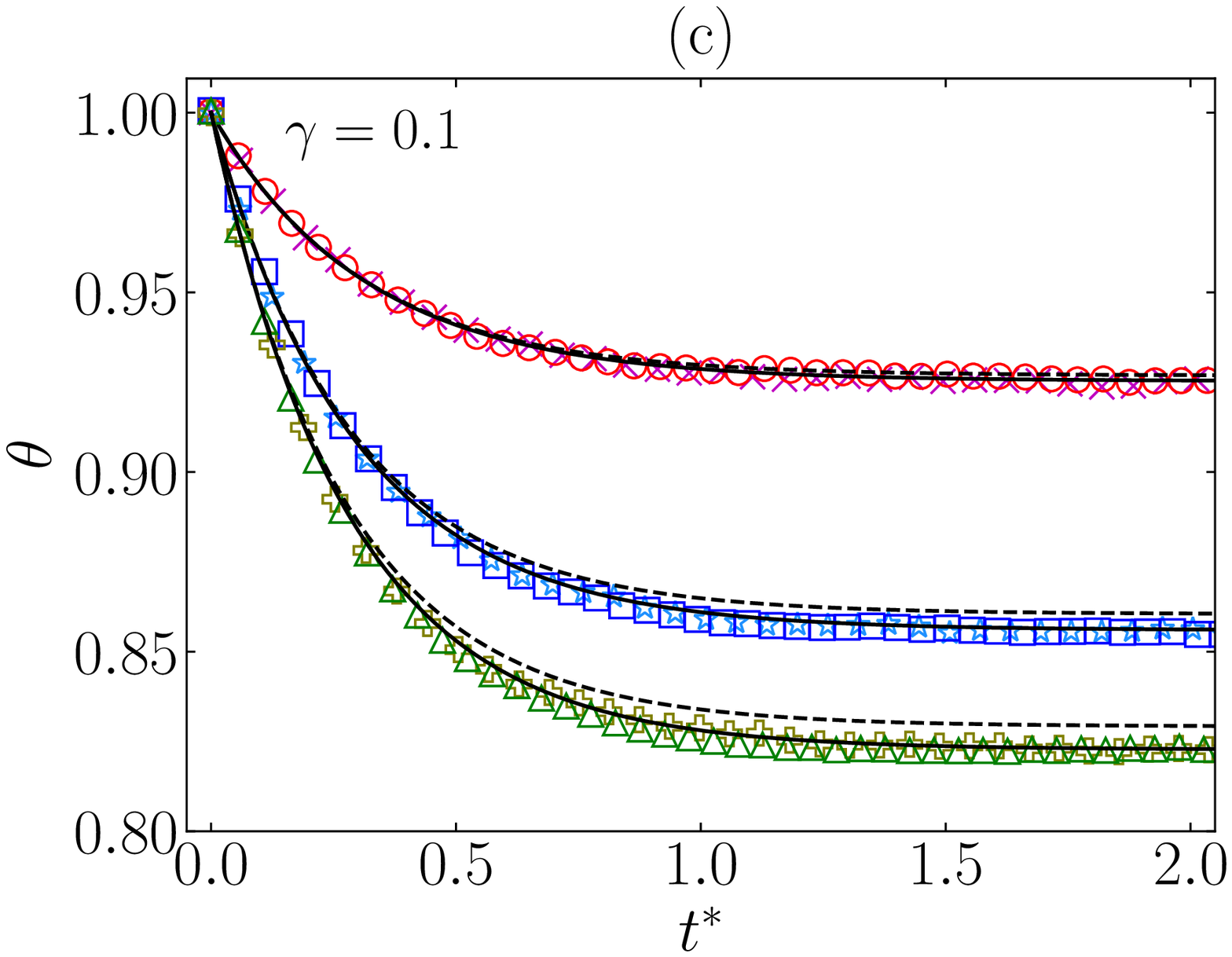}
    \includegraphics[width=0.245\textwidth]{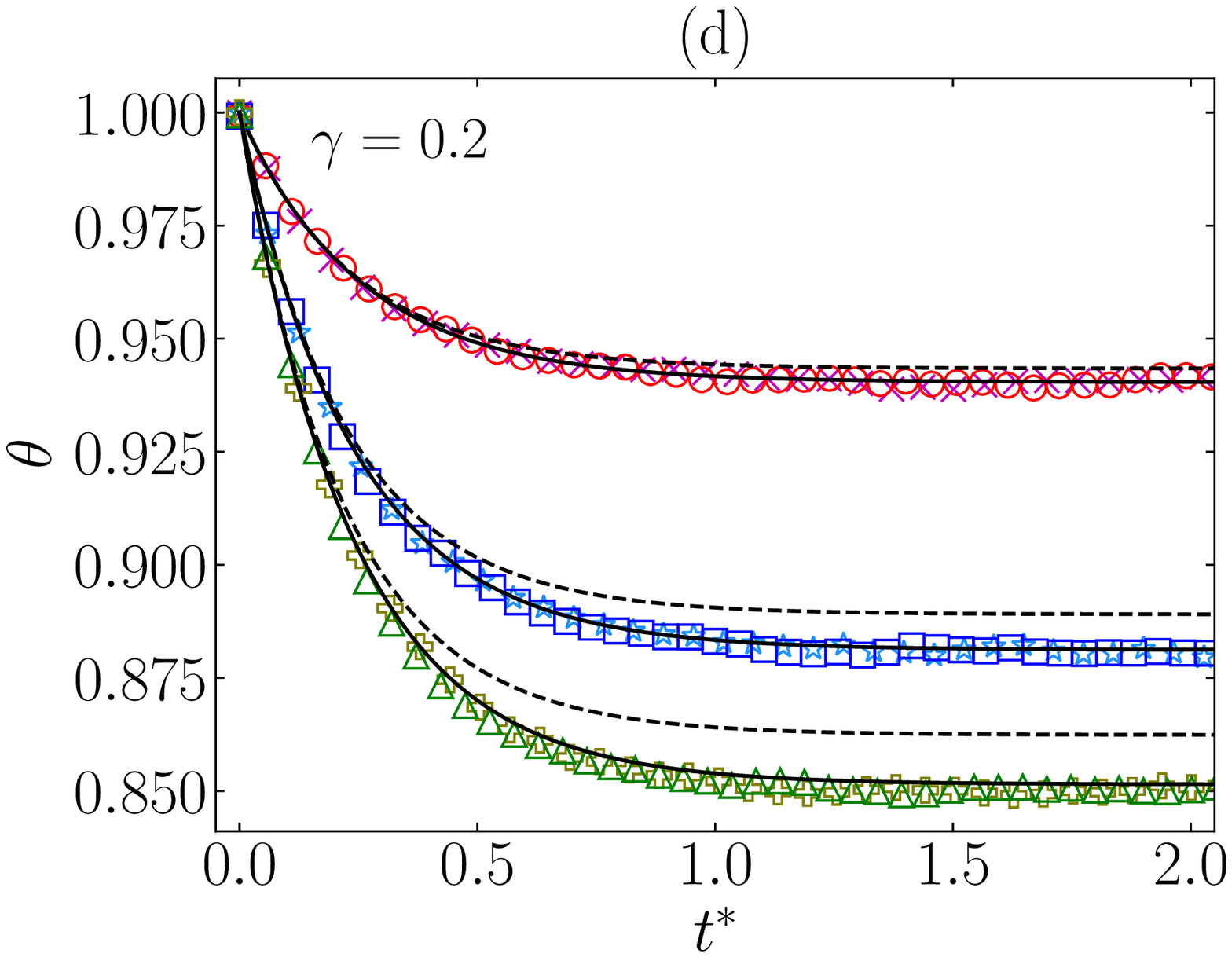}\\
    \includegraphics[width=0.245\textwidth]{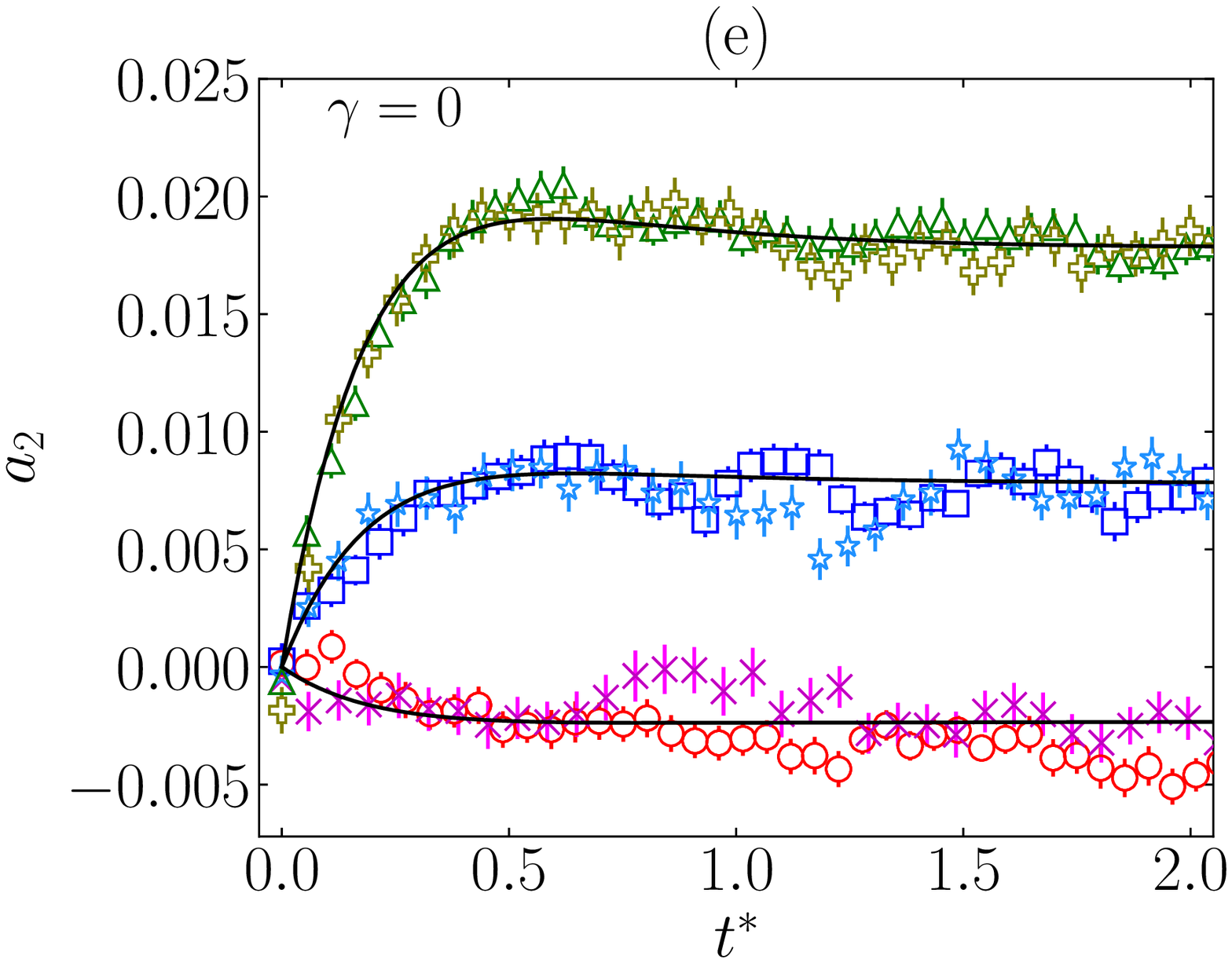}
    \includegraphics[width=0.245\textwidth]{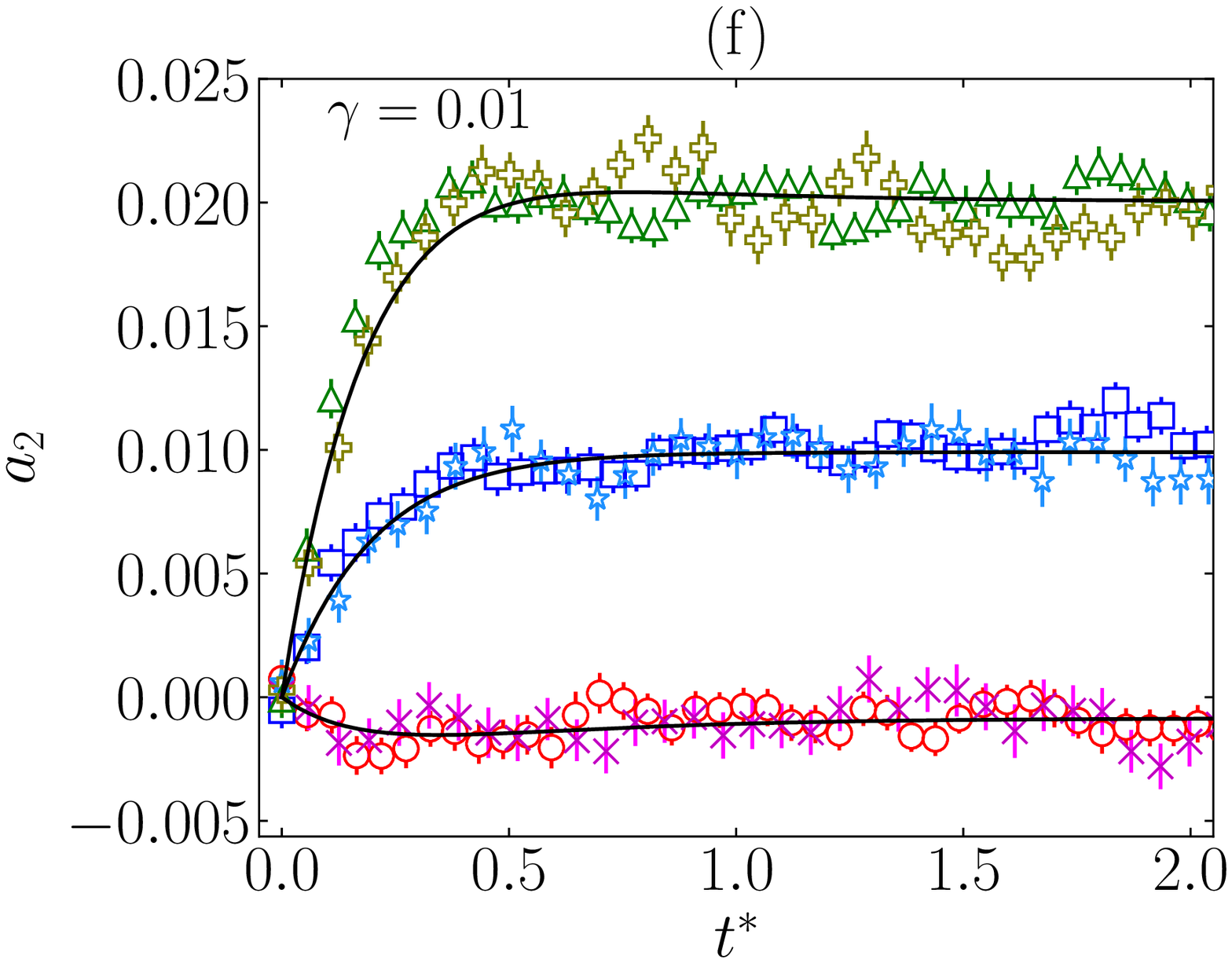}
    \includegraphics[width=0.245\textwidth]{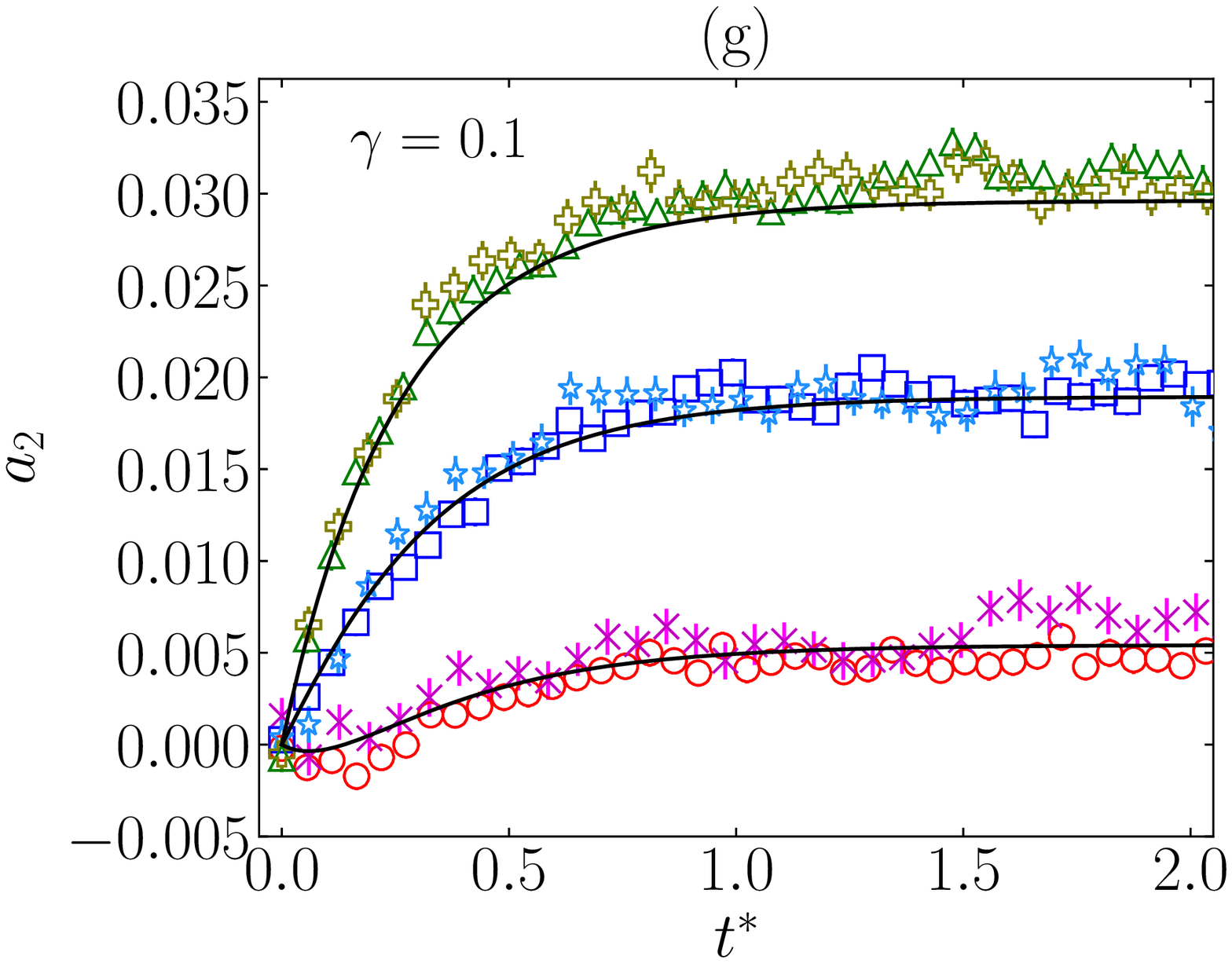}
    \includegraphics[width=0.245\textwidth]{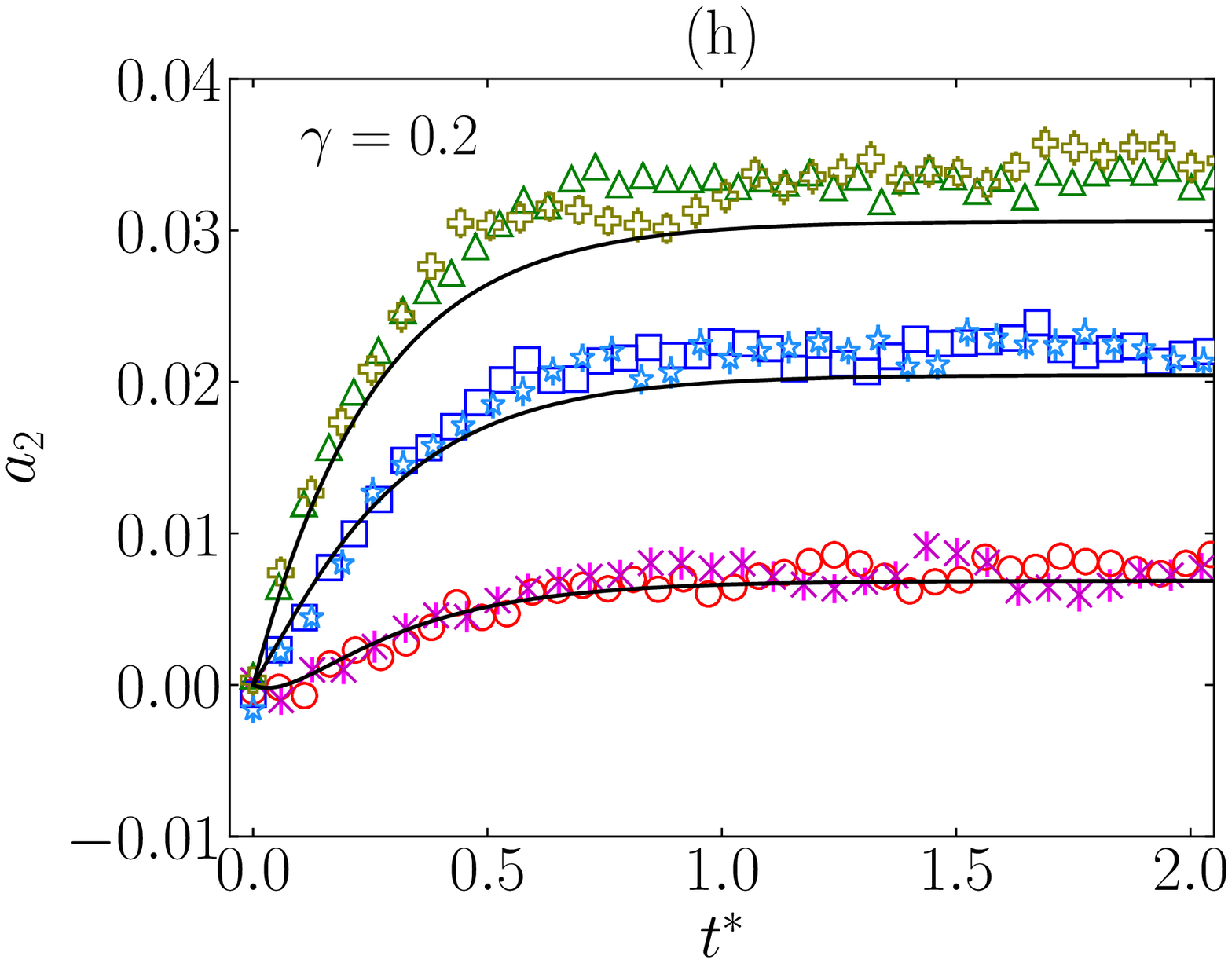}
    \caption{Plots of the time evolution of (\textbf{a}--\textbf{d}) the temperature ratio $\theta(t^*)$ and (\textbf{e}--\textbf{h}) the excess kurtosis $a_2(t^*)$ for $\xi_0^*=1$, $d=3$, and different values of the coefficient of normal restitution ($\alpha=0.8,0.5,0.2$) and the nonlinearity parameter: (\textbf{a},\textbf{e}) $\gamma=0$, (\textbf{b},\textbf{f}) $\gamma=0.01$, (\textbf{c},\textbf{g}) $\gamma=0.1$, and (\textbf{d},\textbf{h}) $\gamma=0.2$. The symbols stand for DSMC ($\circ$, $\square$, $\triangle$) and EDMD ($\times$, $*$, $+$) simulation results, respectively. Dashed (-- --) and solid (----) lines refer to MA (only in panels (\textbf{a}--\textbf{d})) and FSA predictions, respectively. All states are initially prepared with a Maxwellian VDF at the bath temperature, i.e., $\theta^0=1$ and $a_2^0=0$. }
    \label{fig:ev_states_sim}
\end{figure}

Apart from the steady-state values, we have studied the temporal evolution of $\theta$ and $a_2$, starting from a Maxwellian VDF at temperature $T_b$, i.e., $\theta(0)\equiv \theta^0=1$ and $a_2(0)\equiv a_2^0=0$.
Note that this state is that of equilibrium in the case of elastic collisions ($\alpha=1$), regardless of the value of the nonlinearity parameter $\gamma$. The theoretical and simulation results are displayed in Figure~\ref{fig:ev_states_sim} for $d=3$, $\xi_0^*=1$,  and some characteristic values of $\alpha$ and $\gamma$.
We observe that the relaxation of $\theta$ is  accurately predicted by MA, except for the later stage with small $\alpha$ and/or large $\gamma$, in accordance with the discussion of Figure~\ref{fig:steady_states_sim}. This is remedied by FSA, which exhibits an excellent agreement with simulation results in the case of $\theta$ and a fair agreement in the case of $a_2$, again  in accordance with the discussion of Figure~\ref{fig:steady_states_sim}. It is also worth mentioning the good mutual agreement between DSMC and EDMD data, even though fluctuations are much higher in $a_2$ than in $\theta$ because of the rather small values of $|a_2|$.

\subsection{Memory Effects}\label{sec:4.1}

Whereas the temperature relaxation from Maxwellian initial states is generally accurate from MA, it misses the explicit dependence of the temperature evolution on the fourth cumulant (see Equation~\eqref{eq:theta_ev}), which, however, is captured by FSA (see Equation~\eqref{eq:theta_ev_FirstSonine}). This coupling of $\theta$ to $a_2$ is a signal of preparation dependence of the system, hence, a signal of memory effects, as occurs in the elastic case reported in Refs.~\cite{SP20,PSP21,MSP22}.

\subsubsection{Mpemba Effect}

We start the study of memory effects with the Mpemba effect \cite{MO69,BL16,LR17,LVPS17,BKC21}. This counterintuitive phenomenon  refers to situations in which an initially hotter sample (A) of a fluid---or, more generally, a statistical-mechanical system---cools down sooner than an initially colder one (B) in a cooling experiment. We will refer to this as the direct Mepmba effect (DME). Analogously, the inverse Mpemba effect (IME) occurs in heating experiments if the initially colder sample (B) heats up more rapidly than the initially hotter one (A) \cite{LR17,LVPS17,SP20,GKG21,MSP22}. In the special case of a molecular gas (i.e., $\alpha=1$), an extensive study of both DME and IME has recently been carried out \cite{SP20,MSP22}.

\begin{figure}[h!]
    \centering
    \includegraphics[width=0.45\textwidth]{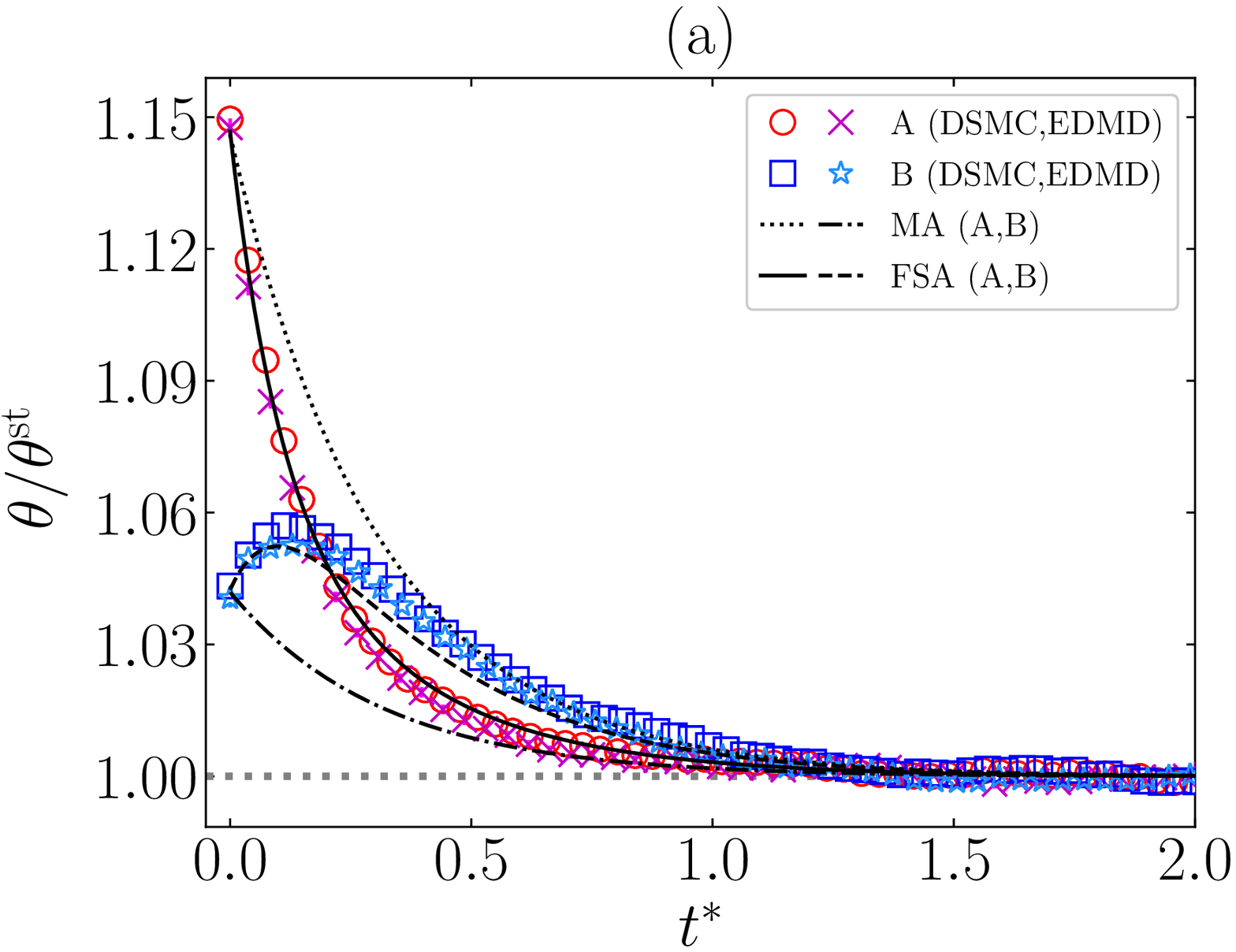}
    \includegraphics[width=0.45\textwidth]{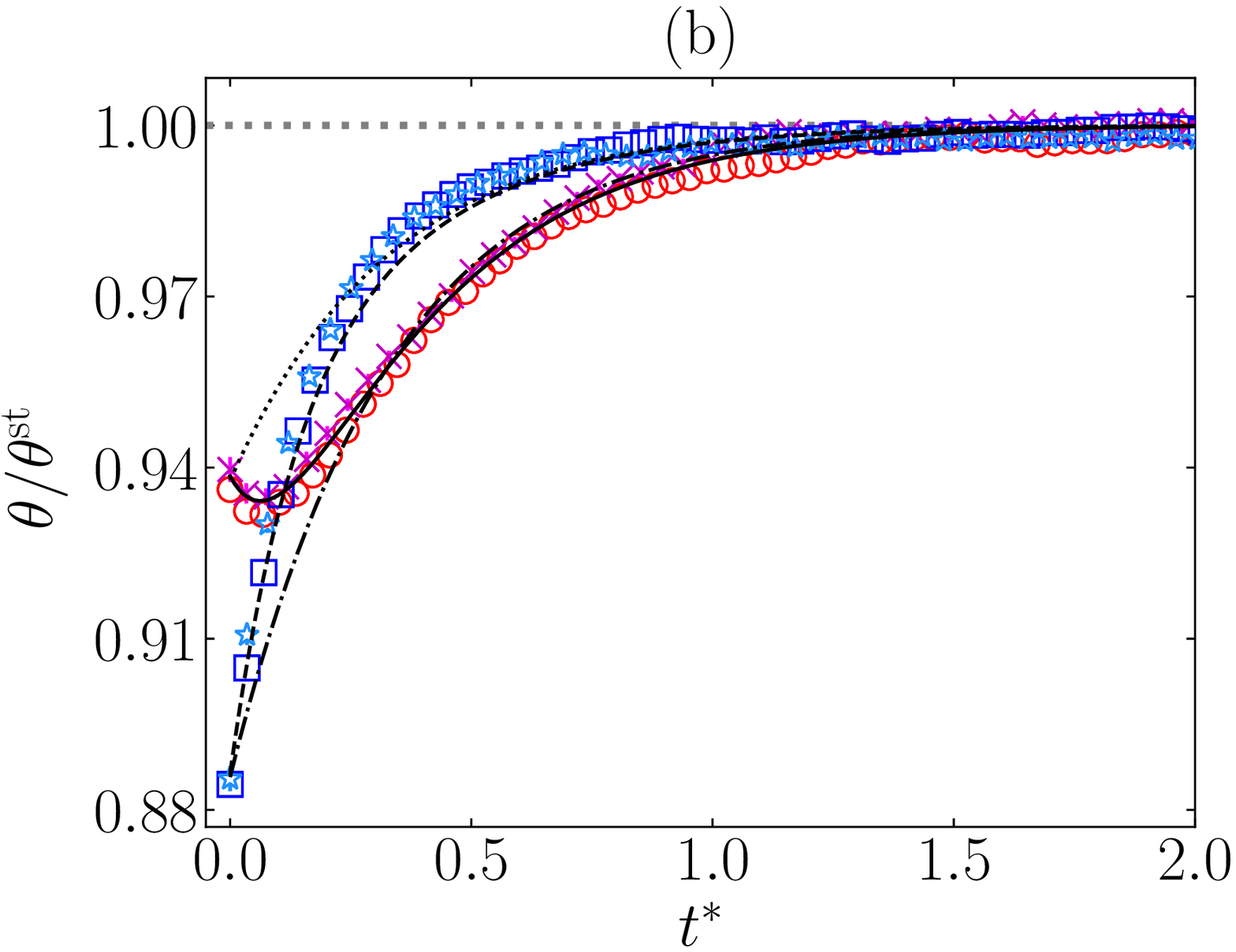}\\
    \includegraphics[width=0.45\textwidth]{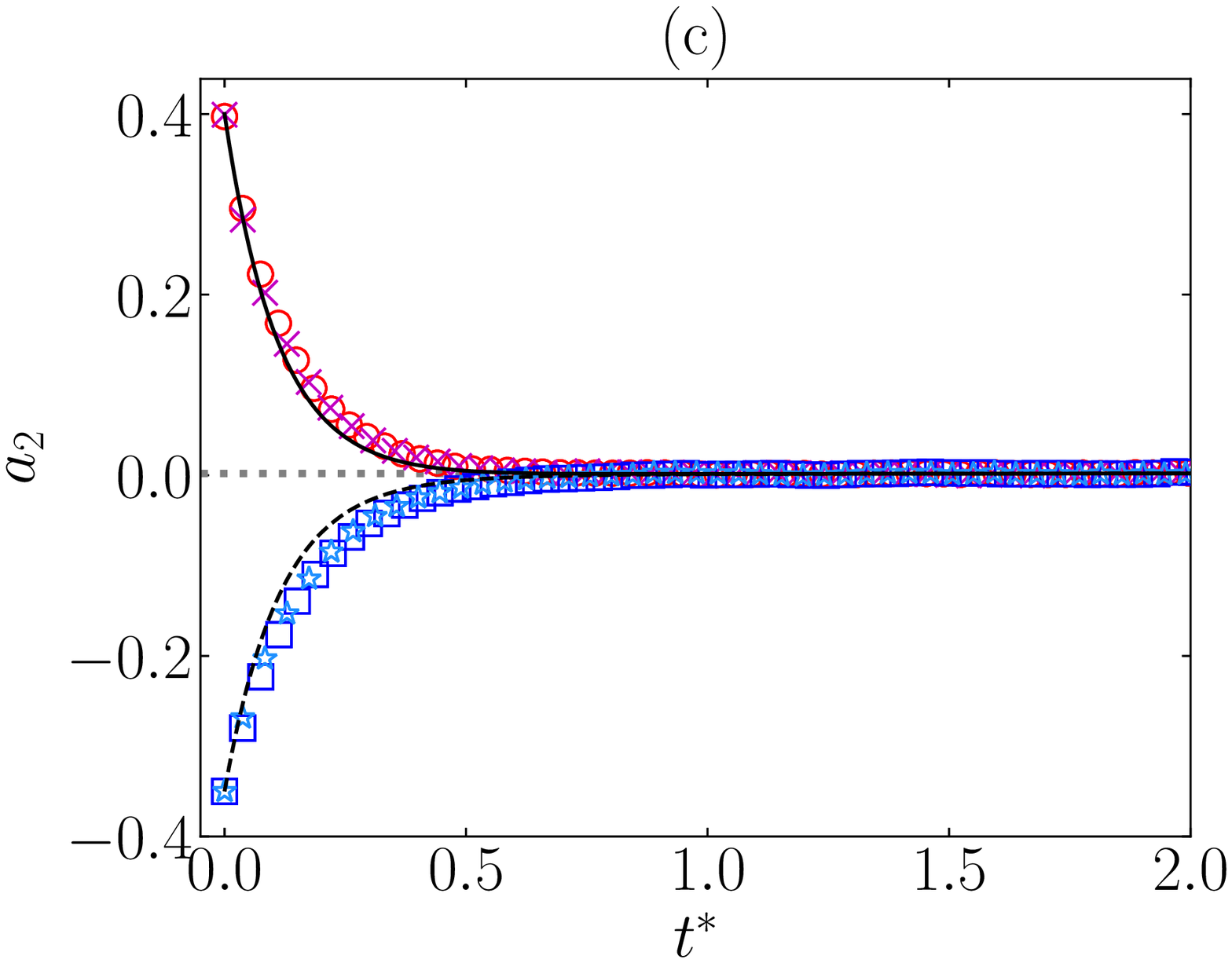}
    \includegraphics[width=0.45\textwidth]{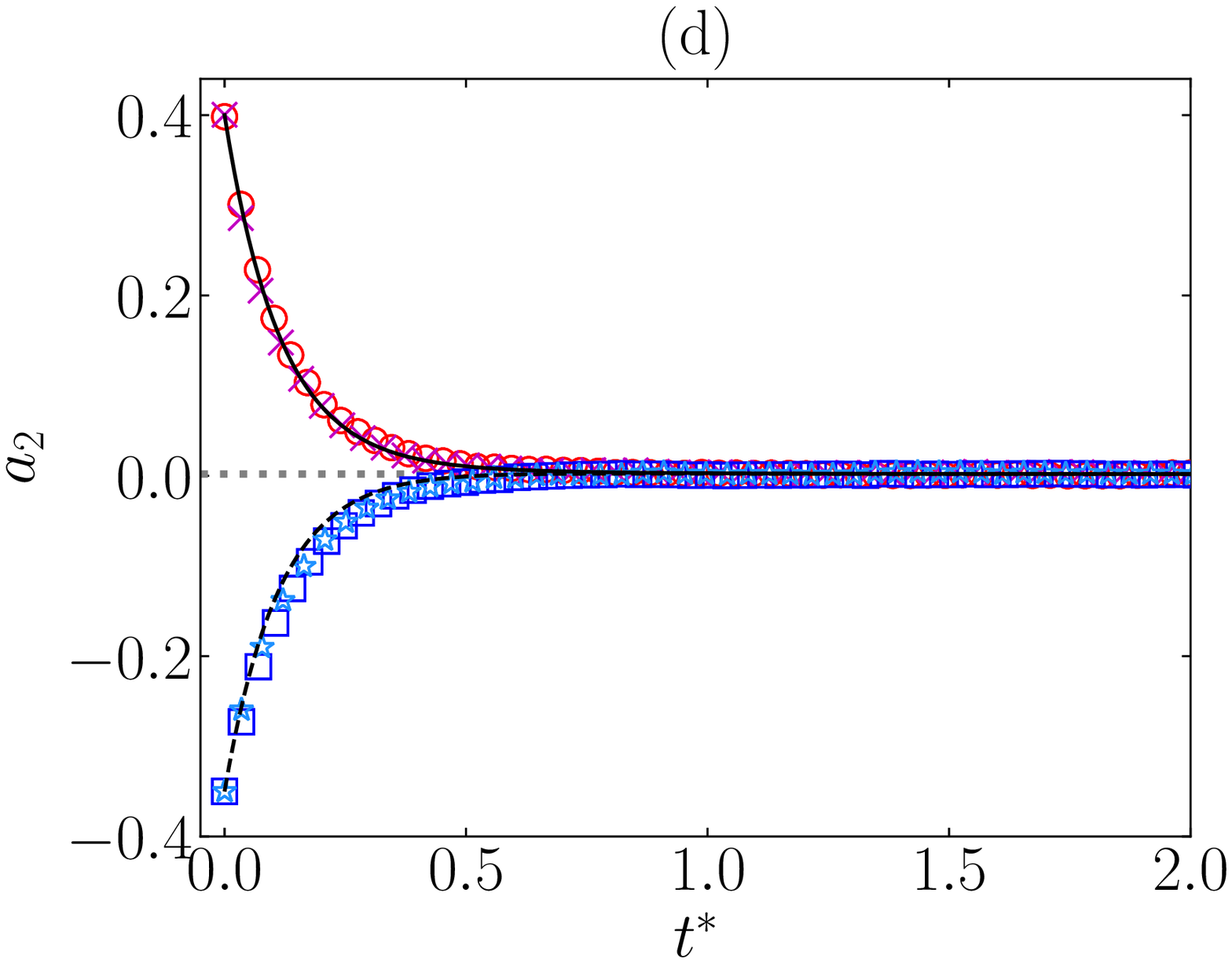}
    \caption{Time evolution of (\textbf{a},\textbf{b}) $\theta(t^*)/\theta^\sta$ and (\textbf{c},\textbf{d})  $a_2(t^*)$ for two samples (A and B) with $\alpha=0.9$, $\xi_0^*=1$, $d=3$, and $\gamma=0.1$. Panels  (a, c) illustrate the DME with initial conditions
$\theta_A^0=1.1\simeq 1.15 \theta^\sta$, $a_{2A}^0=0.4$, $\theta_B^0=1\simeq 1.04\theta^\sta$, $a_{2B}^0=-0.35$, while panels  (b, d) illustrate the IME with initial conditions $\theta_A^0=0.9\simeq 0.94\theta^\sta$,
    $a_{2A}^0=0.4$, $\theta_B^0=0.85\simeq 0.89\theta^\sta$, $a_{2B}^0=-0.35$.
    The symbols stand for DSMC ($\circ$, $\square$) and EDMD ($\times$, $*$) simulation results, respectively. Solid (----) and dashed (-- --) lines correspond to FSA predictions for samples A and B, respectively, whereas black dotted ($\cdots$) and dash-dotted (-- $\cdot$ --) lines in panels (\textbf{a},\textbf{b}) refer to MA predictions for samples A and B, respectively. The gray thin horizontal lines correspond to the steady-state values. Note that $a_2^\sta\neq 0$, despite what panels (\textbf{c},\textbf{d}) seem to indicate because of the vertical scale.    }
    \label{fig:ME_ev}
\end{figure}

Figure~\ref{fig:ME_ev}a,b present an example of DME and IME, respectively. As expected, FSA describes  the evolution and crossing for  temperatures of samples A and B very well. On the contrary, MA does not predict this memory effect. In addition, from Figure~\ref{fig:ME_ev}c,d we can conclude that FSA captures  the relaxation of $a_2$ toward $a_2^\sta\neq 0$ quite well.
\subsubsection{Kovacs Effect}

Next, we turn to another interesting memory effect: the Kovacs effect \cite{K63,KAHR79}. In contrast to the Mpemba effect, the Kovacs effect has a well-defined two-stage protocol and does not involve a comparison between two samples. In the context of our system, the protocol proceeds as follows. First, the granular gas is put in contact with a bath at temperature $T_{b1}$  and initialized at a temperature $T^0>T^\sta_1$, $T^\sta_1=\theta^\sta T_{b1}$ being the corresponding steady-state temperature (note that $\theta^\sta$ is independent of $T_{b1}$ at fixed  $\xi_0^*$). The system is allowed to relax to the steady state during a time window $0<t<t_K$, but then, at $t=t_K$,  the bath temperature is suddenly modified to a new value $T_{b}$, such that $T(t_K)=T^\sta$, $T^\sta=\theta^\sta T_b$ being the new steady-state value. If the system did not retain a memory of its previous history, one would have $T(t)=T^\sta$ for $t>t_K$, and this is, in fact, the result given by the MA. However, the temperature exhibits a hump for $t>t_K$, before relaxing to $T^\sta$.
This hump is a consequence of the dependence of  $\partial_t T$ on the additional variables of the system. According to Equation~\eqref{eq:theta_ev}, and maintained in the FSA, Equation~\eqref{eq:theta_ev_FirstSonine}, the first relevant quantity to be responsible for a possible hump is the excess kurtosis of the VDF, as  occurs in the elastic limit \cite{PSP21}. In fact, at time $t^*=t^*_K$, such that $\theta(t^*_K)=\theta^\sta$, the slope of the temperature according to FSA, Equation~\eqref{eq:theta_ev_FirstSonine}, reads
\begin{equation}
	\dot{\theta}(t^*_K) \approx 2\theta^\sta\left[(d+2)\xi_0^*\gamma\theta^\sta +\frac{\mu_2^{(1)}}{d}\sqrt{\theta^\sta}\right]\left[a_2^\sta -a_2(t^*_K) \right].
\end{equation}
Thus, a nonzero difference $a_2^\sta-a_2(t^*_K)$ implies the existence of a Kovacs-like hump, its sign being determined by that of this difference; that is, we will obtain an upward hump if $a_2(t_K^*)<a_2^\sta$ or a downward hump if $a_2(t_K^*)>a_2^\sta$.

\begin{figure}[h!]
    \includegraphics[width=0.45\textwidth]{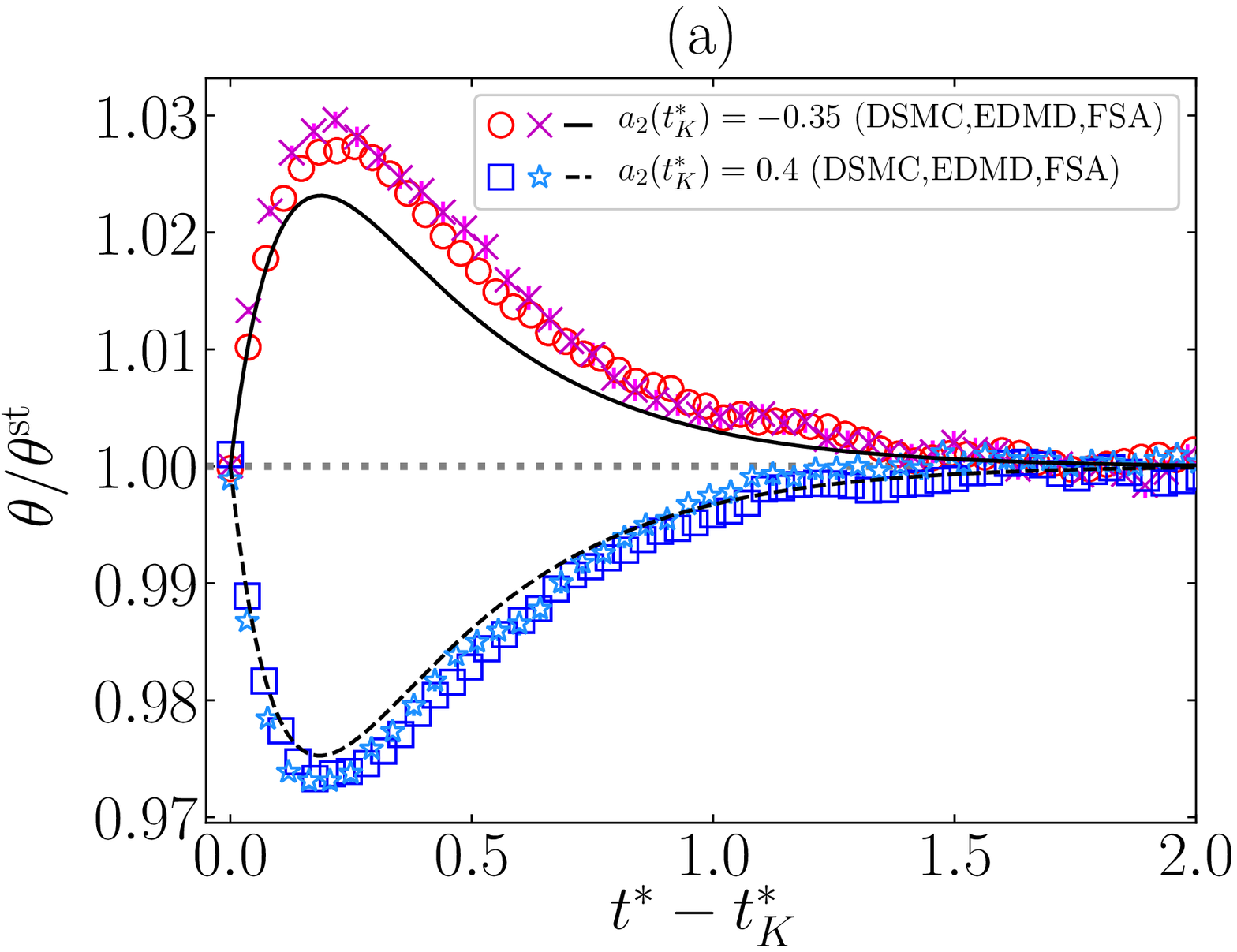}
    \includegraphics[width=0.45\textwidth]{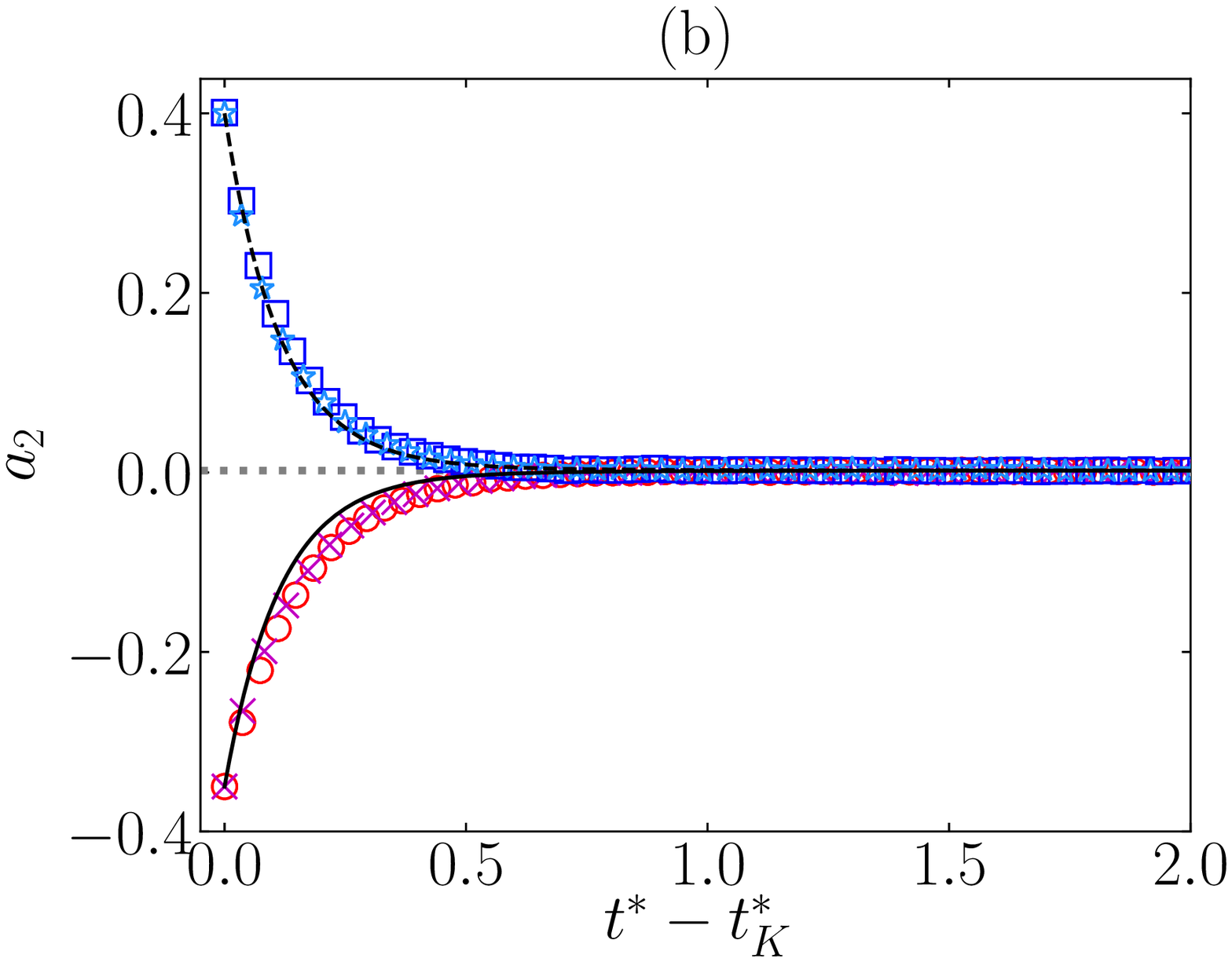}
    \caption{Time evolution for $t^*>t^*_K$ of (\textbf{a}) $\theta(t^*)/\theta^\sta$ and (\textbf{b})  $a_2(t^*)$ for a system with $\alpha=0.9$, $\xi_0^*=1$, $d=3$, and $\gamma=0.1$. The figure  illustrates  Kovacs-like effects with  conditions $\theta(t_K^*)=\theta^\sta$ and either $a_{2}(t_K^*)=-0.35$ ($\circ$, $\square$, ---) or $a_{2}(t_K^*)=0.4)$ ($\times$, $*$, - - -). The symbols stand for DSMC and EDMD  simulation results, while the lines  refer to FSA predictions.}
    \label{fig:Kov_ev}
\end{figure}

For simplicity, in our study of the Kovacs-like effect, we replace the first stage of the protocol ($0<t^*<t_K^*$) by just generating the state at $t^*=t_K^*$ with $\theta(t_k^*)=\theta^\sta$ and $a_2(t_K^*)\neq a_2^\sta$ (see Appendix \ref{sec:ap1}).
The effect is illustrated in Figure~\ref{fig:Kov_ev} for the same system as in Figure~\ref{fig:ME_ev} with the choices $a_2(t_K^*)=-0.35< a_2^\sta$ and $a_2(t_K^*)=0.4> a_2^\sta$. Again, the DSMC and EDMD results agree with each other and with the theoretical predictions. However, in the case $a_2(t_K^*)=-0.35$ (upward hump), Figure~\ref{fig:Kov_ev}a, we observe that the theoretical curve lies below the simulation results. This might be caused by a nonnegligible value of the sixth cumulant $a_{3}(t_K^*)=-0.375$, as reported in Ref.~\cite{MSP22} in the elastic case. Apart from this small discrepancy,  FSA captures the magnitude and sign of the humps, as well as the relaxation of the fourth cumulant, very well.

\section{Conclusions}\label{sec:5}

In this work, we have looked into the dynamics of a dilute granular gas immersed in a thermal bath (at  temperature $T_b$) made of smaller particles but with  masses comparable to those of the grains. To  mathematically characterize  this system, we have worked under the assumptions of Boltzmann's kinetic theory, describing the system by the one-particle VDF, whose evolution is monitored by the EFPE, Equation~\eqref{eq:EFPE}, for the IHS model of hard $d$-spheres. The action of the bath on the dynamics of the granular gas is modeled by a nonlinear drag force and an associated stochastic force. At a given dimensionality $d$, the control parameters of the problem are the coefficient of normal restitution ($\alpha$), the (reduced) drag coefficient at zero velocity ($\xi_0^*$), and the nonlinearity parameter ($\gamma$).

After a general presentation of the kinetic theory description in Section~\ref{sec:2}, we obtained the evolution equation of the reduced temperature $\theta(t^*)\equiv T(t)/T_b$ (Equation~\eqref{eq:theta_ev}), which is coupled explicitly with the excess kurtosis, $a_2$, and depends on every velocity moment through the second collisional moment $\mu_2$ (which is nonzero due to inelasticity). Therefore, the whole dynamics in the context of the EFPE is formally described by Equation~\eqref{eq:theta_ev} and the infinite hierarchy of moment equations given by Equation~\eqref{eq:hierarchy_M}. In order to give predictions, we proposed two approximations. The first one is MA, which consists of assuming a Maxwellian form for the one-particle VDF, whereas the second one, FSA consists of truncating the Sonine expansion of the VDF up to the first nontrivial cumulant $a_2$. Their evolution equations are given by Equations~\eqref{eq:theta_ev_Max} and~\eqref{eq:ev_FSA}, respectively. The predictions for the steady-state values are exposed in Figures~\ref{fig:thetast_th} and~\ref{fig:a2st_th}, which show some small discrepancies in $\theta^\sta$ between MA and FSA as we increase  the inelasticity (decreasing $\alpha$). Moreover, we observed that, for fixed $\alpha$ and $\xi_0^*$, $a_2^\sta$ gets its maximum value when the nonlinearity parameter is  $\gamma=\gamma_{\max}(\alpha,\xi_0^*)$. Another interesting feature is the existence of a critical value $\gamma_c$, such that for $\gamma>\gamma_c$, the values of $a_2^\sta$ are always positive for every value of $\alpha$, while for $\gamma<\gamma_c$, we find $a_2^\sta<0$ for inelasticities small enough. Interestingly, the value of $\gamma_c$ given by Equation~\eqref{eq:gamma_c} is found to be independent of $\xi_0^*$. In addition, some already known limits are recovered in Section~\ref{sec:3.2.2}.

Furthermore, in order to check the predictions from MA and FSA equations, we carried out DSMC and EDMD simulations for hard spheres ($d=3$) with fixed $\xi_0^*=1$ (which corresponds to comparable time scales associated with drag and collisions). First, from Figure~\ref{fig:steady_states_sim}a, we can conclude that, whereas MA provides good predictions of $\theta^\sta$, except for large inelasticities and values of $\gamma$ close to $\gamma_{\max}$, FSA is much more accurate because it takes into account the influence of $a_2^\sta$. The latter approach is generally reliable for $a_2^\sta$, as observed in Figure~\ref{fig:steady_states_sim}b, although, not unexpectedly, it slightly worsens as $|a_2^\sta|$ grows. Relaxation curves starting from a Maxwellian initial state in Figure~\ref{fig:ev_states_sim} show that FSA agrees very well with both DSMC and EDMD; however, MA exhibits good agreement during the first stage of the evolution but becomes less reliable as the steady state is approached.

A relevant feature of these systems, as already studied in the elastic case \cite{SP20,PSP21,MSP22}, is the emergence of memory effects, which are not contemplated by MA. FSA predicts  the emergence of the Mpemba effect very well for both DME and IME, as can be seen in Figure~\ref{fig:ME_ev}. Analogously,  Kovacs-like humps, both upward and downward, are correctly described by FSA, as observed in Figure~\ref{fig:Kov_ev}, although the FSA humps are slightly less pronounced (especially the upward one) than the simulation ones. This is presumably due to the role played by $a_3$ and higher-order cumulants, as occurs in the elastic limit reported in Ref.~\cite{MSP22}.

To conclude, we expect that this work will motivate research about this type of system and the emergence of memory effects. For instance, one can extend the study to other collisional models (such as that of rough spheres), to nonhomogeneous states, or to a more detailed description of the memory effects observed.

\vspace{6pt}

\authorcontributions{
A.M. worked out the approximations and performed the simulations. A.S. supervised  the work.
Both authors participated in the analysis and discussion of the results and worked on the revision and writing of the final manuscript. All authors have read and agreed to the published version of the manuscript.}

\funding{The authors acknowledge financial support from Grant No.~PID2020-112936GB-I00 funded by MCIN/AEI/10.13039/501100011033, and from Grants No.~IB20079 and No.~GR21014 funded by Junta de Extremadura (Spain) and by ERDF
``A way of  making Europe.''
A.M. is grateful to the Spanish Ministerio de Ciencia, Innovaci\'on y Universidades for a predoctoral fellowship FPU2018-3503.}

\acknowledgments{The authors are grateful to the computing facilities of the Instituto de Computaci\'on Cient\'ifica Avanzada of the University
of Extremadura (ICCAEx), where the simulations were run.
\\

\noindent \textbf{Data Availability:} The data presented in this study are available in the online repository \url{https://github.com/amegiasf/GranularNonlinearDrag}}



\conflictsofinterest{The authors declare no conflict of interest. 
}

\abbreviations{
The following abbreviations are used in this manuscript:\\
\noindent 
\begin{tabular}{@{}ll}
DME & Direct Mpemba effect\\
DSMC  & Direct simulation Monte Carlo\\
EDMD & Event-driven molecular dynamics\\
EFPE & Enskog--Fokker--Planck equation\\
FSA & First Sonine approximation\\
IHS & Inelastic hard spheres\\
IME & Inverse Mpemba effect\\
MA & Maxwellian approximation\\
VDF & Velocity distribution function\\
\end{tabular}
}

\appendixtitles{yes} 
\appendix
\section{Simulation Details}\label{sec:ap1}

Throughout the elaboration of this work, we have used two different algorithms to simulate the considered system: DSMC and EDMD methods. Whereas the former is based on statistical properties and subjected to the assumptions of the Boltzmann equation, such as \emph{Stosszahlansatz}, the latter solves the trajectory of each particle without any extra assumption. On the other hand, the original algorithms are slightly modified for the proper collisional model and the interaction with the thermal bath, as explained below.

In general, the simulation results shown in this work are obtained from averaging over $100$ samples in both simulation schemes, and steady-state results come from averaging over $50$ points in the mean trajectory once stationary behavior is observed.

\subsection{Direct Simulation Monte Carlo}\label{subsec:ap1DSMC}

The DSMC algorithm used in this work is based on the original works of G.A. \mbox{Bird~\cite{B94,B13}}, but modified for the IHS collisional model and the implementation of the nonlinear drag. As we considered homogeneous states, only the velocities of the $N$ granular particles, $\left\{\vv_i \right\}_{i=1}^N$, are used to numerically solve  the EFPE. Whereas initial velocities for results in Figures~\ref{fig:steady_states_sim} and~\ref{fig:ev_states_sim} were drawn from a Maxwellian VDF with $\theta^0=1$; in the case of Figures~\ref{fig:ME_ev} and~\ref{fig:Kov_ev}, velocities were initialized from a Gamma VDF (see Refs.~\cite{MS20,MSP22} for additional details). After initialization, particles were updated with a fixed time step, $\Delta t$, much smaller than the mean free time. The method is properly divided into two stages: collision and free streaming \cite{MS00}.

In the collision stage, a number $\lfloor\frac{1}{2}N\omega_{\max}\Delta t\rfloor$ of pairs are randomly chosen with equiprobability---the ignored decimals in the rounding are saved for the next iterative step---$\omega_{\max}$ being an upper bound estimate for the one-particle collision rate. Then, given a chosen pair $ij$, a collision is accepted with probability $\Theta(\vv_{ij}\cdot \ssa_{ij} )\omega_{ij}/\omega_{\max}$, where $\ssa_{ij}$ is a random vector drawn from a uniform probability distribution in the unit $d$-sphere, and $\omega_{ij} = \frac{2\pi^{d/2}}{\Gamma(d/2)} g_c n\sigma^{d-1}|\vv_{ij}\cdot\ssa_{ij}|$. Acceptance implies that the velocities are updated according to the collisional rules in Equation~\eqref{eq:coll_rule}, i.e., $\vv_{i/j}(t)\rightarrow \vv_{i/j}(t+\Delta t)=\vv_{i/j}\pm \frac{1+\alpha}{2}(\vv_{ij}\cdot\ssa_{ij})$.

In the free-streaming stage, each particle velocity is updated according to an Euler numerical algorithm of a Langevin-like equation derived from an It\^{o} interpretation of the Fokker--Planck part of the EFPE (see Ref.~\cite{MSP22}),
\begin{equation}\label{eq:Langevin_vel}
	\vv_i(t)\rightarrow \vv_i(t+\Delta t) = \vv_i(t) - \left[\xi(v_i(t))-2\xi_0\gamma \right]\vv_i \Delta t+\chi(v_i(t))\sqrt{\Delta t}\mathbf{Y}_i,
\end{equation}
where $\mathbf{Y}_i$ is a random vector drawn from a Gaussian probability distribution with unit variance, $P(\mathbf{Y})= (2\pi)^{-d/2} e^{-Y^2/2}$.

In the implementations of the DSMC algorithm, we used $N=10^4$ hard spheres ($d=3$) and a time step $\Delta t=10^{-2}\lambda/v_b$, $\lambda= (\sqrt{2}\pi n\sigma^2)^{-1}$ being the mean free path.

\subsection{Event-Driven Molecular Dynamics}\label{subsec:ap1EDMD}

EDMD methods compute the evolution of particles driven by
events: particle--particle collisions, boundary effects, or other more complex interactions. Analogously  to the splitting described in the DSMC algorithm, free streaming of particles occurs between two consecutive events. Here, we need to consider the influence of the stochastic and drag forces not only in the velocities but also in the positions of the $N$ granular particles, $\left\{\mathbf{r}_i\right\}_{i=1}^N$. In order to account for this, we followed the \emph{approximate Green Function} algorithm proposed in Ref.~\cite{S12}. Whereas the velocities are updated according to Equation~\eqref{eq:Langevin_vel}, the positions follow
\begin{equation}\label{eq:Langevin_pos}
	\mathbf{r_i}(t)\rightarrow \mathbf{r_i}(t+\Delta t)= \mathbf{r}_i(t)+\vv_i(t)\Delta t\left[1-\Delta t\frac{\xi(v_i(t))-2\gamma\xi_0}{2} \right]+\frac{1}{2}\chi(v_i(t))\Delta t^{3/2} \mathbf{W}_i,
\end{equation}
where $\mathbf{W}_i=\mathbf{Y}_i+\sqrt{5/3}\mathbf{Y}'_i$, $\mathbf{Y}^\prime_i$ being another random vector drawn from $P(\mathbf{Y})= (2\pi)^{-d/2} e^{-Y^2/2}$.

In the EDMD simulations, we defined a set of $N=8\times10^3$ hard spheres ($d=3$), with a reduced number density $n\sigma^3=10^{-3}$, implying a box length $L/\sigma=2\times10^2$, and used a time step $\Delta t\approx 10^{-3}\lambda/v_b$. Periodic boundary conditions were imposed, and no inhomogeneities were observed.


\reftitle{References}





\end{document}